\documentclass[twocolumn,
 amsmath,amssymb,
 pra,floatfix
]{revtex4-1}

\usepackage{graphicx}
\usepackage{dcolumn}
\usepackage{bm}
\usepackage{epstopdf}

\begin{document}

\title{Modulational instability and frequency combs in WGM microresonators with backscattering}

\author{Nikita M. Kondratiev}
 \email{noxobar@mail.ru}
\author{Valery E. Lobanov}%
\affiliation{%
 Russian Quantum Center, 143025 Skolkovo, Russia
}%

\date{\today}

\begin{abstract}
We introduce the first principle model describing frequency comb generation in a WGM microresonator with the backscattering-induced coupling between the counter-propagating waves. {Elaborated model provides deep insight and accurate description of the complex dynamics of nonlinear processes in such systems.} We analyse the backscattering impact on the splitting and reshaping of the nonlinear resonances,  demonstrate  backscattering-induced modulational instability in the normal dispersion regime and subsequent frequency comb generation. We present and discuss novel features of the soliton comb dynamics induced by the backward wave.
\end{abstract}

\maketitle


\section{Introduction}
Compactness, high quality factors and energy efficiency of the optical whispering gallery mode (WGM) microresonators make a promise for a variety of scientific and technological applications of these devices \cite{Matsko2006,Strekalov_2016,Lin:17}.
Significant breakthroughs in this area were the discoveries of the Kerr frequency combs (or microcombs) \cite{Kippenberg-2007, PASQUAZI20181,Gaeta2019} and of the associated dissipative Kerr solitons (DKS) in  microresonators \cite{herr2014temporal,Kippenbergeaan8083}. More recently DKS generation and combs have been demonstrated using a variety of compact semiconductor-based sources. In particular, coupling of a high-quality-factor (high-Q) microresonator to a diode laser has been demonstrated to provide laser stabilization and linewidth reduction via the self-injection locking effect \cite{VASSILIEV1998305, Kondratiev:17,Galiev:18,photonics5040043,Sprenger:09,Liang:10}.
Usually, narrow-linewidth laser sources have been used for microresonator pumping and frequency comb generation. However, recently the generation of DKS was demonstrated with the laser diode operating in the self-injection locking regime with crystalline \cite{Pavlov_18np} and on-chip \cite{Raja2019} microresonator. The self-injection locking effect appears due to the Rayleigh scattering inside the microresonator \cite{Gorodetsky:00} when subsequent backward wave provides resonant feedback that can result in a significant reduction of the laser linewidth. The described backward wave also interacts nonlinearly with the forward wave and may influence frequency comb generation and dynamics.
For example the appearance of the modulational instability induced by the cross-phase modulation was shown  for co-propagating waves \cite{Agrawal1987,ZHANG2005193,Tanemura:03,Li:19}. Also frequency comb generation at normal group velocity dispersion (GVD) was {demonstrated in case of the coupling between different co-propagating spatial or polarizational mode families existing simultaneously in the microresonators} \cite{Jang:16,Xue-2015,Ramelow:14}. 
Recently a number of results on the impact of the linear and nonlinear couplings between the counter-propagating waves has been reported \cite{PhysRevLett.99.173603,Yoshiki:15} including generation of DKS \cite{Fujii:17,Vahala_18np,Joshi:18}. 
However, the absence of the consistency and transparent justification of the applied models call for a revision of this problem in the view of its persistent importance for practical applications. Here we are starting from the first principles and derive a generic and numerically tractable model describing interaction between the counter-propagating waves that includes a space reversal effect in the coupling term and the terms accounting for the opposite signs of the group velocities. The latter complicates numerical approaches to the problem, since it requires tracing of the two well separated oppositely propagating pulses along the ring circumference.  However, we demonstrate that under the quite generic conditions these terms, as well as the nonlinear cross coupling, can be averaged out. Our extensive numerical studies of both generic and averaged models demonstrate an excellent agreement between the two for high finesse systems {and allow us to find the bound, where this approximation is applicable. For low finesse systems (like fiber ring resonators} \cite{Nielsen2019}{) the full equations should be used. The other main focus is the problems of DKS and modulational instability in the normal GVD regime and the associated frequency combs generation}.

\section{Equation Derivation}
We start the analysis of the nonlinear processes in high-Q WGM microresonators with backscattering from wave equation for the electric field $\vec E$ with a nonlinear term restricted to the Kerr nonlinearity \cite{boyd2013nonlinear}:
\begin{equation}
\label{MaxwellEquation}
\nabla\times\nabla \times \vec E+\frac{\hat\epsilon}{c^2}\frac{\partial^2 \vec E}{\partial t^2}=-\frac{\chi_3}{ c^2}\frac{\partial^2 (\vec E |\vec E|^2) }{\partial t^2}.
\end{equation}
{We consider the case of the lumped pump}, when the coupler is spatially separated and the coupling region is localized (evanescent coupling, like prism, tapered fiber or waveguide). Then we introduce the field as a sum of the {WGM field $\vec E_{\rm w}$ and the pump field $\vec E_{\rm p}$} and the permittivity $\hat\epsilon=1+\hat\chi_{\rm w}+\hat\chi_{\rm c}$, {where the $\hat\chi_{\rm w}$ is susceptibility of the resonator and $\hat\chi_{\rm c}$ -- susceptibility of the coupler, which are nonzero only in the corresponding regions to reflect the geometry under consideration.}  The pump field is small compared to the WGM field (Q-factor enhanced) in the WGM region, so the nonlinear terms with it can be neglected. As we represent one unknown field with two unknowns, we should impose the restriction or separate equation \eqref{MaxwellEquation} into two. Thus, we collect the terms referred to the microresonator into the first equation {and the others to the second}. Then we get the microresonator field equation with the pump term similar to \cite{Gorodetsky:99} and coupler equation as follows{:}
\begin{align}
\label{MasterEquation}
&\nabla\!\!\times\!\!\nabla\!\!\times\!\!\vec E_{\rm w}+\frac{\hat\epsilon}{c^2}\frac{\partial^2 \vec E_{\rm w}}{\partial t^2}=-\frac{\chi_3}{ c^2}\frac{\partial^2 (\vec E_{\rm w} | E_{\rm w}  |^2) }{\partial t^2}-\frac{1}{\epsilon_0 c^2}\frac{\partial^2 \vec P_{p}}{\partial t^2},\nonumber\\
&\nabla\!\!\times\!\!\nabla\!\! \times\!\! \vec E_{\rm p}+\frac{1+\chi_c}{c^2}\frac{\partial^2 \vec E_{\rm p}}{\partial t^2}=0,
\end{align}
where $\vec P_p=\chi_{\rm w}\vec E_p$ is the polarization induced by the pump field. We do not consider the coupler equation here, but assume that it can be solved so that the solution can be represented in the form $\vec P_p=\epsilon_0\chi_{\rm w} \Re\left(F\vec f_p(r)e^{-i\omega t}\right)$, where $\vec f_p(r)$ is the pump field profile, $F$ is the pump {electrical field amplitude and $\Re$ stands for the real part operator}. 
The pump frequency $\omega$ is close to some microresonator mode with the number $m_0$ {(here we use a single letter for the triplet of modal indices for simplicity and will expand to the full notation later)}. Further we expand the electric field {of the microresonator} in terms of the forward and backward spatial modes of the microresonator $\vec e^+_\mu(r)$ and $\vec e^-_\mu(r)$ ($\mu$ is the modal number offset from $m_0$), oscillating with the pump frequency $\omega$: 
 \begin{equation}
 \label{solform}
 \vec E_{\rm w}(r, t)=\Re\!\sum^{\mu=N}_{\mu=-N}\!\!\! \left(A_\mu(t) \vec e^{+}_\mu(r)+B_\mu(t) \vec e^{-}_\mu(r)\right)e^{-i\omega t},
 \end{equation}
{with $A_\mu$ and $B_\mu$ being the complex forward and backward propagating field amplitudes}. Here we do not consider the problem of the modes orthogonality and eigenfrequency complexity related to the openness of the system \cite{Deych2011,Hill1990}. Similar to \cite{Gorodetsky:99} { we just assume that they are solutions of the microresonator equation of the system}  \eqref{MasterEquation} with zero right-hand side, no coupler, no roughness and the following orthogonality relation satisfied $\int_{V_w} \vec e_\mu^\dag(r)\hat \epsilon\vec e_\nu(r) d^3r=n^2V_{\mu}\delta_{\mu\nu}$ for the finite volume $V_w$ close to the volume of the microresonator, where $n$ is the refractive index of the microresonator material, $V_\mu$ is the effective mode volume, and the eigenfrequencies of interest $\omega_\mu$ are purely real. The losses, including coupler-related ones, will be introduced into the equation {for the resonator electric field from} \eqref{MasterEquation} manually. To introduce the backward wave generation, we should represent the permittivity as a sum of the main ideal part and its perturbation due to the surface roughness and/or the pump coupler $\hat\epsilon\rightarrow\hat\epsilon+\delta\epsilon$ \cite{Gorodetsky:00}. {We also note that this perturbation consists of regular and random parts as the latter is what makes the backscattering nonzero}.

For further analysis, we substitute \eqref{solform} into \eqref{MasterEquation} and use the slowly varying amplitude approach. Assuming that $\dot{A_\mu}\ll A_\mu\omega$, we write out the parts of \eqref{MasterEquation}, removing fast-oscillating in time terms with $2\omega$ time exponents. 
Removing $\Re$ from combined equations, we use the orthogonality of the modes to separate the forward and backward wave equations. To simplify the overlap integrals, we use the cylindrical symmetry of the WGM problem and extract the azimuthal dependence  as  $\vec e_{\mu}^\pm(\vec r)=\vec e^{\pm}_{p,q,\mu+m_0}(r,z) e^{\pm i(\mu+m_0)\varphi}$. Here, $p$ and $q$ are the transverse mode numbers that were previously implicit in $m_0$ and $\mu$, while $\mu$ is now the offsets of the azimuthal indices from $m_0$. For large enough $m_0$ (which is usually the case), the transverse  profiles of all the comb modes can be assumed to be similar and independent of $\nu$. At the same time, we assume that the modes are orthogonal over the $p$ and $q$ indices, so that we get nonzero results only inside the mode family (fixed $p$ and $q$) and omit their indices for the sake of compactness.
Performing azimuthal integration, most of the terms zero out and we get the coupled mode equation system (CMES) \cite{Chembo2010,Cherenkov:17}. At this point the loss terms $\kappa A_\mu$ and $\kappa B_\mu$ {are} added {($\kappa$ is the loaded linewidth of the pumped mode)} and the time is normalized to the {photon lifetime} $\tau=t\kappa/2$. {The transition to the microresonator free spectral range (FSR) grid is made and the field amplitude is made dimensionless} so that $A_\mu\rightarrow \sqrt{\frac{4n^2\kappa}{3\chi_3\omega}} a_\mu e^{-i\mu D_1t}$, $B_\mu\rightarrow \sqrt{\frac{4n^2\kappa}{3\chi_3\omega}}b_\mu e^{-i\mu D_1t}$ and we get

\begin{widetext}
\begin{align}
\dot{a}_\mu=&-(1+i\alpha_\mu) a_\mu+i\sum_{\nu}(\delta\alpha^+_{\mu\nu}a_{\nu}+\beta_{\mu\nu}b_{\nu})e^{-id_1(\nu-\mu)\tau} 
+i\hspace{-1em}\sum_{\mu'=\nu+\eta-\mu}\hspace{-1em}\Theta^{^+\nu\eta}_{\mu'\mu} a_{\nu}a_{\eta}a^{*}_{\mu'}+i2\hspace{-1em}\sum_{\mu'=\mu-\nu+\eta}\hspace{-1em}\Theta^{'\nu\eta}_{\mu'\mu} 
a_{\nu}b_{\eta}b^{*}_{\mu'}e^{-i2(\nu-\mu)d_1\tau}+\nonumber\\ &+f_\mu e^{id_1\mu\tau},\nonumber\\
\label{normalizedB}
\dot{b}_\mu=&-(1+i\alpha_\mu) b_\mu+i\sum_{\nu}(\delta\alpha^-_{\mu\nu}b_{\nu}+\beta_{\mu\nu}^\dag a_\nu)e^{-id_1(\nu-\mu)\tau} 
+i\hspace{-1em}\sum_{\mu'=\nu+\eta-\mu}\hspace{-1em}\Theta^{^-\nu\eta}_{\mu'\mu} b_{\nu}b_{\eta}b^{*}_{\mu'}+i2\hspace{-1em}\sum_{\mu'=\mu+\nu-\eta}\hspace{-1em}\Theta^{'\nu\eta}_{\mu\mu'}
b_{\eta}a_{\nu}a^{*}_{\mu'}e^{-i2(\eta-\mu)d_1\tau}.  
\end{align}
\end{widetext}
Here 
$\beta_{\mu \nu}=\frac{Q}{2n^2}\int \vec e^{+*}_{\mu}\delta \epsilon \vec e^-_{\nu}\frac{dV}{V_{\mu}}$ {is the coefficient of the linear coupling of the $\mu$-th forward mode and the $\nu$-th backward mode (forward-backward wave coupling or backscattering coefficient)} \cite{Gorodetsky:00},
$\delta\alpha^\pm_{\mu \nu}=\frac{Q}{2n^2}\int \vec e^{\pm*}_{\mu}\delta \epsilon \vec e^{\pm}_{\nu}\frac{dV}{V_{\mu}}$ is the normalized frequency deviation due to the roughness and presence of the coupler,
$Q=\omega/\kappa$ is the microresonator loaded quality factor,
$d_1/2=D_1/\kappa$ is the microresonator finesse ($D_1$ is the microresonator FSR),
$\alpha_\mu=\frac{2}{\kappa}(\omega_\mu-\mu D_1-\omega)$ is the normalized pump frequency detuning,
$\omega_\mu$ is the $\mu$-th mode eigenfrequency,
$f_\mu=i \frac{ \omega}{2n^2}\chi_{\rm w}\sqrt{\frac{3\chi_3\omega}{n^2\kappa^3}}F\int\vec e^{+*}_{\mu}\vec f\frac{dV}{V_{\mu}}$ is the normalized pump amplitude term and the fourth order direct and cross-term transverse overlap integrals
\begin{align}
\label{overlapint}
\Theta&^{^\pm{\nu\eta}}_{\mu'\mu}=\frac{1}{3}\int(\vec e^\pm_{\nu}\vec e^\pm_\eta)(\vec e^{\pm*}_{\mu'}\vec e^{\pm*}_{\mu})+2(\vec e^{\pm*}_{\mu'}\vec e^{\pm}_{\eta})(\vec e^{\pm}_{\nu}\vec e^{\pm*}_{\mu}) \frac{rdrdz}{V_{\mu}},\nonumber\\
\Theta&^{' {\nu\eta}}_{\mu'\mu}=\frac{2}{3}\int 
(\vec e^{+}_\nu\vec e^{-}_\eta)(\vec e^{-*}_{\mu'}\vec e^{+*}_\mu) +
(\vec e^{-*}_{\mu'}\vec e^{-}_\eta)(\vec e^{+}_\nu\vec e^{+*}_\mu)+\nonumber\\
&+
(\vec e^{-*}_{\mu'}\vec e^{+}_\nu)(\vec e^{-}_\eta\vec e^{+*}_\mu)
\frac{rdrdz}{V_{\mu}}.
\end{align}
Note that here we already used the azimuthal exponent orthogonality to reduce the summation in \eqref{normalizedB} and only transverse surface integration is left in \eqref{overlapint}. Due to the orthogonality relation of the modes, for small anisotropy, we can estimate the first transverse integral $\Theta^{^\pm\nu\eta}_{\mu'\mu}\approx1$. The cross-term integral can also be assumed $\Theta^{'\nu\eta}_{\mu\mu'}\approx1$ for the modes with the same polarization and close to 1/3 for different polarizations. Note also that for the crossed-polarization case one should consider different eigenfrequencies $\omega'_\mu$ (and corresponding detunings and dispersion coefficients) for the backward waves. In this work we assume the same polarization of the forward and the backward waves. 

{The equation system} \eqref{normalizedB} {is bulky, but simple in structure. Both equations consist of the common resonance term, the mode shift and linear mode coupling term (the first sum), self-phase modulation and cross phase modulation terms (the second and third sum respectively). They are also
similar to the standard equation for Kerr soliton comb generation} \cite{herr2014temporal}, except the coupling and the nonlinear cross-action terms. 
It was shown in \cite{Gorodetsky:99}, that $i \frac{ \omega}{2n^2}\chi_{\rm w}\int\vec e^{+*}_{\mu}\vec f\frac{dV}{V_{\mu}}=\sqrt{\eta_\mu\kappa D_1}$, where $\eta_\mu\in[0;1]$ is the pump coupling coefficient, so that $\eta_\mu\kappa$ is the mode decay rate related to the presence of the pump coupler.
Rewriting $F$ in terms of the input power, we get $f_\mu=\sqrt{\frac{6\chi_3Q\eta_0 P_{\rm input}}{\kappa n^4\epsilon_0 V_0}}\sqrt{\frac{nS}{n_cS_c}}\delta(\mu)$, where $S$ and $S_c$ are beam areas in the WGM and the coupler, $n_c$ -- coupler refractive index and $\delta(\mu)$ is close to the delta-function and appears due to the phase matching conditions with the coupler. This expression for the pump term coincides with the commonly used one \cite{herr2014temporal,Kippenbergeaan8083} when the beam areas and refractive indices are close, which is usually the case.

If the cavity finesse is large enough ($d_1/2\gg\alpha_\mu$), the linear forward-backward coupling terms and the nonlinear cross-action terms contain fast-oscillating components that have no practical influence on the system dynamics. Thus, the summations can be truncated to $i(\delta\alpha^+_{\mu\mu}a_{\mu}+\beta_{\mu\mu}b_{\mu})$ and $i2a_{\mu}\sum|b_{\eta}|^2$ in Eqs. \ref{normalizedB}.
For simplicity we include the $\delta\alpha_{\mu\mu}$ into $\alpha_\mu$.
{In} \cite{Gorodetsky:00} {it was shown that the backscattering coefficient $\beta_{\mu\mu}$ is dependent on the azimuthal number, but this dependence is negligible near large pumped mode number $m_0$ in bulk microresonators for the number of comb lines up to 200. So in this work we assume that $\beta_{\mu\mu}=\beta$ is independent on azimuthal number. Such approximation is not good for integrated microresonators, where $\beta_{\mu\mu}$ was found to exhibit strong random variations over $\mu$ in the same mode family} \cite{Zhu2010,Li:12}. {However this has no impact on the stationary solutions and linear stability analysis that is presented in the following and can be taken into account in numerical modelling using appropriate $\beta_{\mu\mu}$ in later works.}
The pump term is simultaneously reduced to $f\delta_{\mu0}$. Note, that this approximation means that we account only for the linear and nonlinear coupling of the forward and backward modes having the same modal indices. 
Introducing notations $\widetilde x_m=\sum_{\nu=0}^{N-1}x_\nu e^{-2\pi i\nu m/N}$ for the discrete Fourier transform (dft) and  $\widehat x_\nu=\frac{1}{N}\sum_{p=0}^{N-1}x_p e^{2\pi i\nu p/N}$ for the inversed one (idft), where $N$ is the number of modes, and calculating the triple sums as described in \cite{HANSSON2014134}, we get the following equations:
\begin{align}
\dot{a}_\mu=&-(1+i\alpha_\mu) a_\mu+i\beta b_{\mu}+i \widehat{\widetilde a\widetilde a\widetilde a^*}_\mu +i2a_{\mu}\sum|b_{\eta}|^2 +f \delta_{\mu 0},\nonumber\\  
\label{finalBsimp}
\dot{b}_\mu=&-(1+i\alpha_\mu) b_\mu+i\beta^* a_\mu+i \widehat{\widetilde b\widetilde b\widetilde b^*}_\mu+i2b_{\mu}\sum|a_{\eta}|^2.  
\end{align}

Note, that for the accelerated calculation of the cross-action terms for the full equations \eqref{normalizedB}, we use idft for $b$ instead of dft.  It can also be shown in this case that for the term in the forward wave, with current definition of dft normalization a factor of $N^2$ will appear.

\begin{figure*}[ht]
\centering
\includegraphics[width=0.3\linewidth]{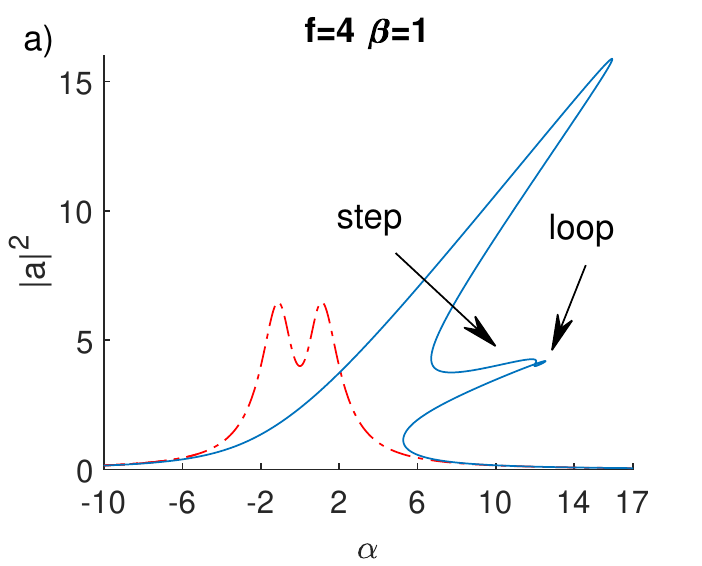}
\includegraphics[width=0.3\linewidth]{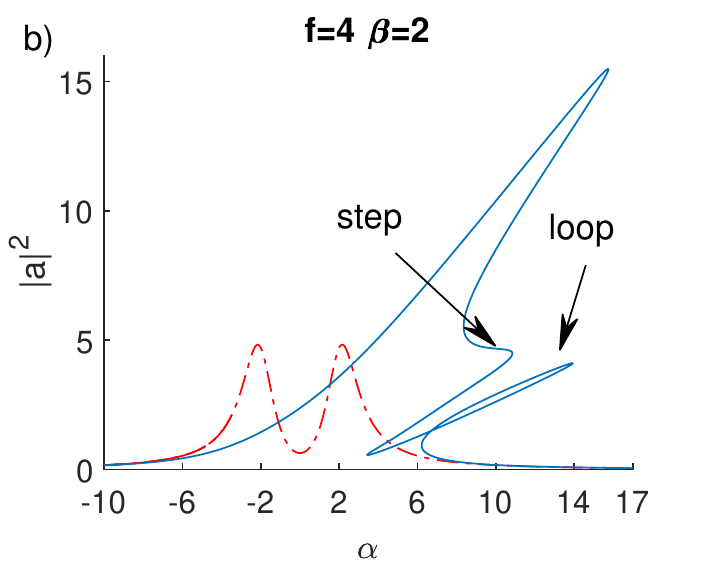}
\includegraphics[width=0.3\linewidth]{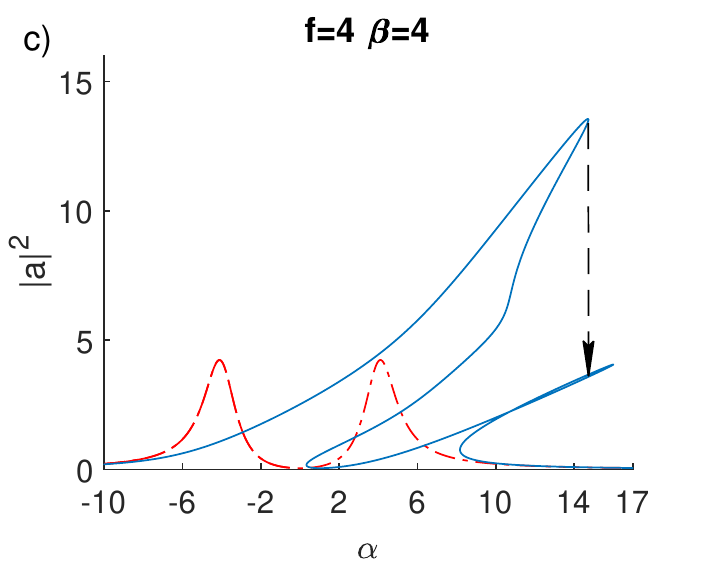}
\caption{Forward wave linear {(red dash-dotted line)} and nonlinear {(blue solid line)} resonance curves for different $\beta$ at $f=4$. Dashed arrow in the right panel shows that the loop crosses the main resonance region {and the system will jump from the first to the second branch while tuning the frequency. Plotted variables are dimensionless.}}
\label{fig:stationary2}
\end{figure*}

Usually, the Lugiato-Lefever type equation (LLE), widely used for modeling of comb generation processes \cite{PhysRevLett.58.2209,PhysRevA.87.053852,doi:10.1098/rsta.2018.0113}, is got from the CMES \eqref{normalizedB} before the normalization with substitution $A(\varphi)=\sum A_\mu e^{i\mu\varphi}$,  $B(\varphi)=\sum B_\mu e^{-i\mu\varphi}$ and  $\omega_\mu=\omega_0+D_1\mu+D_2\mu^2$, where $D_2$ is the GVD coefficient.
Note, that we choose the minus sign at exponent in the expression for $B$ to emphasize that it rotates in the opposite direction. 
For the sake of brevity, we again include the frequency deviation due to the roughness and presence of the coupler into $\omega_\mu$.
\begin{align}
\dot{A}=&-i(\omega_0-\omega-i\kappa/2)A-D_1\frac{\partial A}{\partial\varphi}+iD_2\frac{\partial^2 A}{\partial\varphi^2}+F(\varphi)+\nonumber\\
&+igA(|A|^2+2|B|^2)+i\int s(\varphi,\theta)B(\theta)\frac{d\theta}{2\pi},\nonumber\\ 
\label{LLEB0}
\dot{B}=&-i(\omega_0-\omega-i\kappa/2)B+D_1\frac{\partial B}{\partial\varphi}+iD_2\frac{\partial^2 B}{\partial\varphi^2}+\nonumber\\
&+igB(|B|^2+2|A|^2)+i\int s^*(\theta,\varphi)A(\theta)\frac{d\theta}{2\pi},
\end{align}
where $s(\varphi,\theta)=\sum e^{i\nu\theta}\frac{\kappa}{2}\beta_{\nu\mu}e^{i\mu\varphi}$. 
Assuming that linear coupling occurs for the forward and backward modes with the same indices ($\beta_{\nu\mu}=\beta\delta_{\nu\mu}$), that corresponds also to the high-finesse case, we get the coupling terms in the form $isB(-\varphi)$ and $is^*A(-\varphi)$. 
Note that in case $\beta_{\nu\mu}=\beta$, meaning that all modes couple equally, we get them in the form  $isB(0)\delta(\varphi)$ and $is^*A(0)\delta(\varphi)$. In this article only the first case is considered. 
Here we also note that the LLE approach looks less convenient for the non-trivial coupling case.

However there is a problem in numerical modeling as the common substitution $\varphi'=\varphi+D_1t$, $A(\varphi'-D_1t,t)=A'(\varphi',t)$ and $B(\varphi'-D_1t,t)=B'(\varphi',t)$ does not remove the fast-rotating term with $D_1$ from the second equation. 
So we derive LLE directly from the simplified equations \eqref{finalBsimp} with $a(\varphi)=\sum a_\mu e^{i\mu\varphi}$ and $b(\varphi)=\sum b_\mu e^{-i\mu\varphi}$. 
\begin{align}
\dot{a}=&-(1+i\alpha)a+id_2\frac{\partial^2a}{\partial\varphi^2}+i\beta b(-\varphi)+ia(|a|^2+2P_b)+f,\nonumber\\ 
\label{LLEBsimp}
\dot{b}=&-(1+i\alpha)b+id_2\frac{\partial^2b}{\partial\varphi^2}+i\beta^*a(-\varphi)+ib(|b|^2+2P_a),
\end{align}
where $d_2=2D_2/\kappa$, $\alpha=\alpha_0$ and $P_a=\sum |a_\mu|^2=\int |a(\varphi)|^2\frac{d\varphi}{2\pi}$ and  $P_b=\sum |b_\mu|^2=\int |b(\varphi)|^2\frac{d\varphi}{2\pi}$ are the average intensities. {The appearance of the averaged intensities instead of local ones in cross-action terms reflect the fact that the fields perform fast rotation in opposite directions and average each other. It can be also shown that the sign in the linear interaction term argument is a consequence of the cylindrical symmetry.} Note, that similar equations were used in \cite{Fujii:17,Vahala_18np}, but the signs in the argument of the linear coupling term are different and nonlinear cross-action terms are absent or depend on the local intensity values instead of the averaged values. Our calculations show that these differences may affect the boundaries of soliton existence and stability domains at anomalous GVD.

\section{Stationary solutions and Linear Stability Analysis}
To investigate the frequency comb generation process in such system more accurately, we use the linear stability analysis (LSA) approach. First, we study the homogeneous solutions of the stationary form of \eqref{finalBsimp} {for the pumped mode ($\mu=0$)}
\begin{align}
\label{stationary}
0=&-\left(1+i\alpha\right)a_0+i\beta b_0+i a_0(|a_0|^2+2|b_0|^2)+if,\nonumber\\
0=&-\left(1+i\alpha\right)b_0+i\beta a_0+i b_0(|b_0|^2+2|a_0|^2).
\end{align}
It is well-known, that linear coupling between the counter-propagating waves splits each of the cavity resonances \cite{Weiss:95,PhysRevLett.99.173603,Gorodetsky:00,Kippenberg:02} (see the {(red dash-dotted line)} lines in Fig. \ref{fig:stationary2}). Solving \eqref{stationary} numerically, we found that in a nonlinear system this splitting happens in a different fashion. At weak coupling (small values of $\beta$), a characteristic step appears on the resonance curve (see Fig. \ref{fig:stationary2}{a}). Then, as the linear coupling coefficient increases, a loop is formed at the tip of the step.  With further increase of the coupling coefficient, loop goes down (see Fig. \ref{fig:stationary2}{b}), becomes separated from the step and, finally, splits off. With a further growth of the coupling parameter $\beta$, the step disappears, and the loop turns into a second, narrower resonance (see Fig. \ref{fig:stationary2}{c}). 

The characteristic values of $\beta$, at which the resonance curve transformations occur, depend on the pump intensity (they are collected in the Table \ref{charPoint}). Note, that $f=0.2$ corresponds to almost linear behavior and $f=1.241$ -- to multi-stability appearance. Our further investigations show that these transformations of the nonlinear resonance curve highly affect the process of comb generation.

\begin{table}[ht]
\begin{ruledtabular}
\caption{\label{charPoint} The characteristic values of $\beta$. {The rows "step" and "loop" correspond to the coupling coefficient values at which respective curve feature formation starts; "split" is the value of $\beta$ at which the loop becomes separated from the main resonance; "cross" is the  value of $\beta$ at which the loop of the resonance curve comes out from bellow the main resonance region and thus can be reached after the first branch is over} (see Fig.\ref{fig:stationary2}, right panel). Presented variables are dimensionless.}
\begin{center}
\begin{tabular}{ccccccccc}
f     & 1.5  &   2 & 2.2  & 2.5  &  3   &   4  &  5   &  6   \\ine
step &0.4  & 0.4 & 0.4  & 0.4  & 0.4  & 0.35 & 0.3  & 0.28 \\
loop  & -  &  -  & 1.07 & 0.94 & 0.93 & 0.96 & 0.98 & 0.99 \\
split &  -  &  -  & 0.84 & 0.87 & 1.11 & 1.39 & 1.6  & 1.7  \\
cross & 0.4 & 0.85 & 1.02 & 1.3  & 1.85 & 3.25 & 5.1  & 7.25 \\
\end{tabular}
\end{center}
\end{ruledtabular}
\end{table}

Before proceeding to the LSA we note, that full and simplified equations have the same homogeneous solutions.
Then we analyze stability of the full system \eqref{LLEB0}
using anzats
\begin{align}
\label{anzatsexp1}
&a=a_0+a_1\exp(im\phi)+a_2^*\exp(-im\phi),\nonumber\\
&b=b_0+b_1\exp(-im\phi)+b_2^*\exp(im\phi),
\end{align}
where $a_0$ and $b_0$ are stationary homogeneous solutions of \eqref{stationary} and $a_{1,2}$, $b_{1,2}$ are perturbations.
After the substitution \eqref{anzatsexp1} into \eqref{LLEB0}, small terms neglection and separation according to the azimuthal exponents we get
\begin{align}
&i\dot a_1  =(p_m-i+d_1m)a_1  -a_0^2a_2  -2a_0(b_0b_1^*+b_0^*b_2^*)-\beta b_1, \nonumber\\
&i\dot a_2^*=(p_m-i-d_1m)a_2^*-a_0^2a_1^*-2a_0(b_0^*b_1+b_0b_2)-\beta  b_2^* ,\nonumber\\
&i\dot b_1  =(p_m-i+d_1m)b_1  -b_0^2b_2  -2b_0(a_0a_1^*+a_0^*a_2^*)-\beta^*a_1, \nonumber\\
&i\dot b_2^*=(p_m-i-d_1m)b_2^*-b_0^2b_1^* -2b_0(a_0^*a_1+a_0a_2)-\beta^*a_2^* ,
\label{eq1}
\end{align}
where $p_m=d_2m^2+\alpha-2(|a_0|^2+|b_0|^2)$. {We add complex conjugated equations to close up the system self-consistently,} introduce $(a_1,a_1^*,a_2,a_2^*,b_1,b_1^*,b_2,b_2^*)^T=\vec xe^{\lambda t}$ and derive the eigenvalues problem for the instability growth rate $\lambda$
\begin{align}
\label{8matrix}
&(\lambda+1)\vec x=\hat M_8\vec x,
\end{align}
where $\hat M_8$ is 8x8 matrix.

\begin{figure*}[t]
\centering
\includegraphics[width=0.45\linewidth]{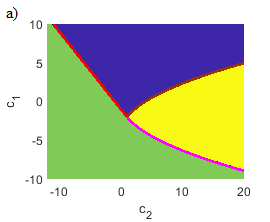}
\includegraphics[width=0.47\linewidth]{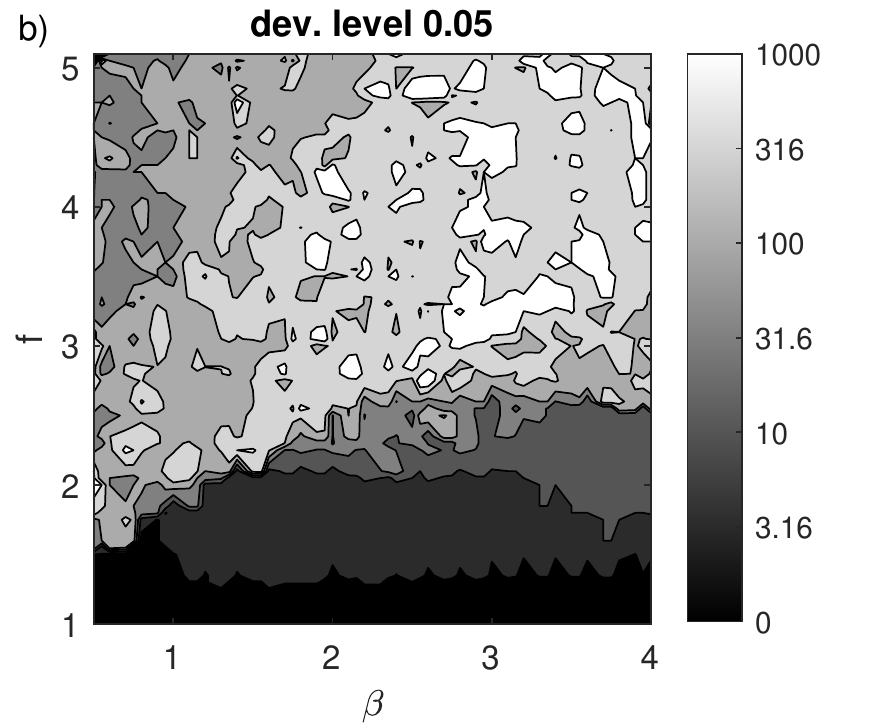}
\caption{{a)} The root map of \eqref{charEqSimp}. The {blue (dark)} area provides stable solutions, {yellow (light) and green (gray)} -- oscillatory and non-oscillatory unstable ones. The {red (black)} line corresponds to $c_1+c_2=-1$, {brown (dark gray)} -- $c_2=(c_1+4)^2/4$ and {magenta (gray)} -- $c_2=c_1^2/4$, critical point is (1;-2). 
{b)} The $d_1$ values, for which the deviation between maximal real parts of eigenvalues is less than $5$\%. Plotted variables are dimensionless.}
\label{fig:charEqSimp}
\end{figure*}

We should remark here that $m=0$ represents a special case resulting in a simpler 4x4 eigenvalue problem.
To build the matrix for the reduced system \eqref{LLEBsimp} we use that $\int \exp(\pm im\theta)d\theta/(2\pi)=\delta_{m,0}$. So, for the reduced system and $m\neq0$ we get the equations \eqref{eq1} without the $d_1$-terms and the cross-terms ($a_0b_0$-terms and conjugation combinations). Eventually, the stability matrix in this case is only 4x4. So, for reduced system we get the characteristic equation
\begin{align}
\label{charEqSimp}
(\lambda+1)^4+c_1(\lambda+1)^2+c_2=0,
\end{align}
\begin{align}
c_1=&2p_m^2-P_{a0}^2-P_{b0}^2+2|\beta|^2+4R_{ab}\delta_{m,0},\nonumber\\
c_2=&(p_m^2-P_{a0}P_{b0}-|\beta|^2)^2-(P_{a0}-P_{b0})^2p_m^2-R_{ab}+\nonumber\\
&-4\delta_{m,0}(R_{ab}+4P_{a0}P_{b0})(P_{b0}+p_m)(P_{a0}+p_m)+\nonumber\\
&4\delta_{m,0}(R_{ab}^2+R_{ab}|\beta|^2),
\end{align}
and $R_{ab}=2\Re[a_0^*b_0\beta]$, $P_{a0}=|a_0|^2$, $P_{b0}=|b_0|^2$. The stability map for this equation is shown in Fig. \ref{fig:charEqSimp}. This {allows} us to highlight the instability regions at each resonance curve in Figs. \ref{fig:dyn_anom}, \ref{fig:dyn_norm}. Basically, the step is stable until the loop forms at its tip. Another stable region usually appear between the loop and the main curve, when they become separated. This happens at slightly higher $\beta$ then the "split" event from Table \ref{charPoint}.

Now we compare the results of LSA obtained from the full system of equations \eqref{normalizedB} and the simplified high-finesse equations \eqref{finalBsimp} (and thus \eqref{LLEB0} and \eqref{LLEBsimp}). We solve \eqref{8matrix} for both 8x8 matrix of the full system for different $d_1$ and for the high-finesse case \eqref{charEqSimp} and compare the roots with maximum real parts. The right panel of Fig. \ref{fig:charEqSimp} shows the $\beta$-$f$ map of $d_1$ where the difference is less then $5$\%. This means that for each combination of the pump amplitude $f$ and the coupling coefficient $\beta$, the solutions of the full and reduced problem are very close if the finesse value $d_1/2$ exceeds the value indicated in the map. 

\section{Numerical modeling}
\begin{figure}[ht]
\centering
\includegraphics[width=0.47\linewidth]{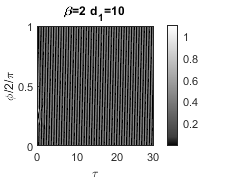}
\includegraphics[width=0.47\linewidth]{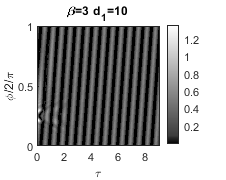}
\includegraphics[width=0.47\linewidth]{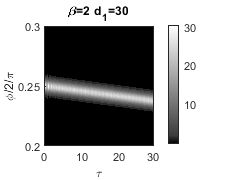}
\includegraphics[width=0.47\linewidth]{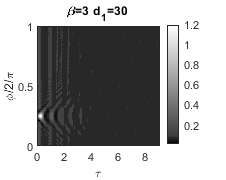}
\includegraphics[width=0.47\linewidth]{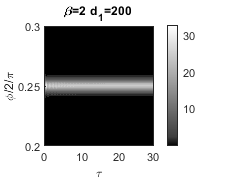}
\includegraphics[width=0.47\linewidth]{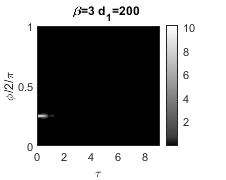}
\caption{Evolution of the initial soliton for different linear coupling coefficients $\beta$ and finesse values $d_1/2$ at fixed $\alpha=12$, $f=4.11$ and $d_2=0.01$. The bottom figures coincide with the high-finesse case. Plotted variables are dimensionless.}
\label{fig:finesse}
\end{figure}

To  check more accurately the applicability of the discussed model simplification we perform direct modelling of the soliton propagation with the full equations \eqref{normalizedB} for the different finesse values. Analysing different solutions of the full system, it is found that for the reasonable values of $\beta$ and $f$ the solution converges to that of the simplified high-finesse equations \eqref{finalBsimp} if the finesse value exceeds some critical value, depending on $\beta$ and $f$.
The Figure \ref{fig:finesse} shows the propagation of the soliton
\begin{align}
a_{\rm sol}(\varphi)=\sqrt{2\alpha}{\rm \, sech\,}\sqrt{\frac{\alpha}{d_2}}(\varphi-\pi/2)
\end{align}
for different $d_1$, $\beta$ and fixed $\alpha=12$, $f=4.11$ and $d_2=0.01$. 

\begin{figure*}[t]
\centering
\includegraphics[width=0.47\linewidth]{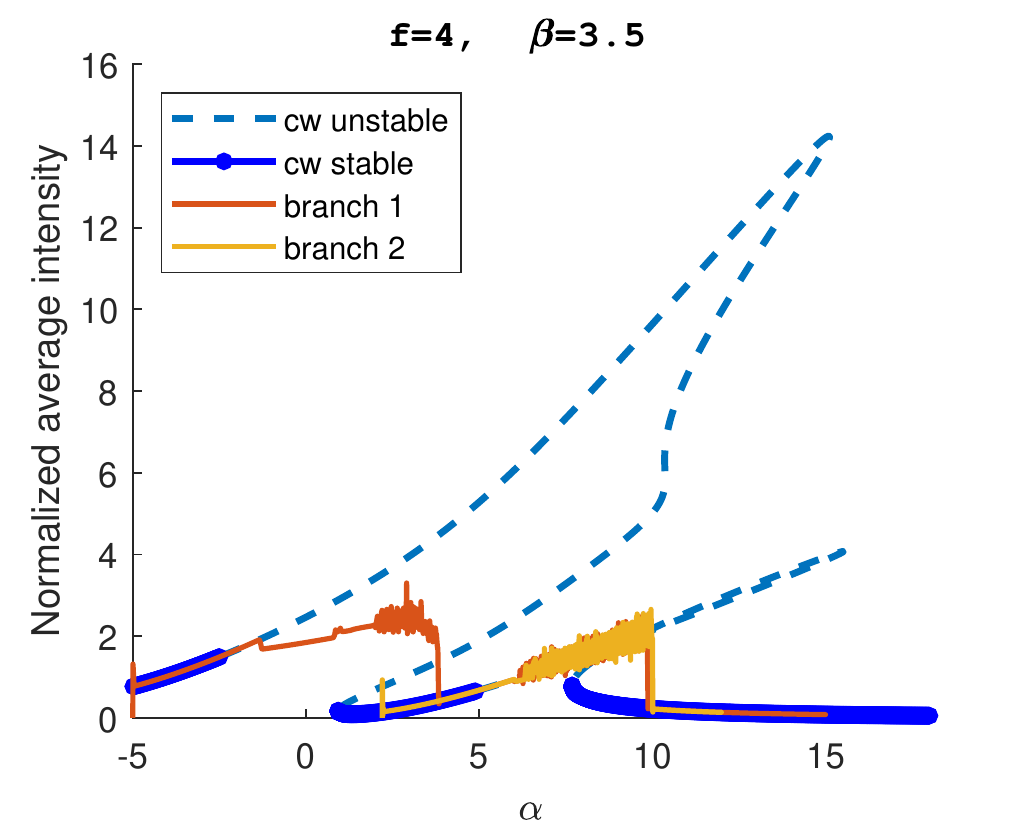}
\includegraphics[width=0.47\linewidth]{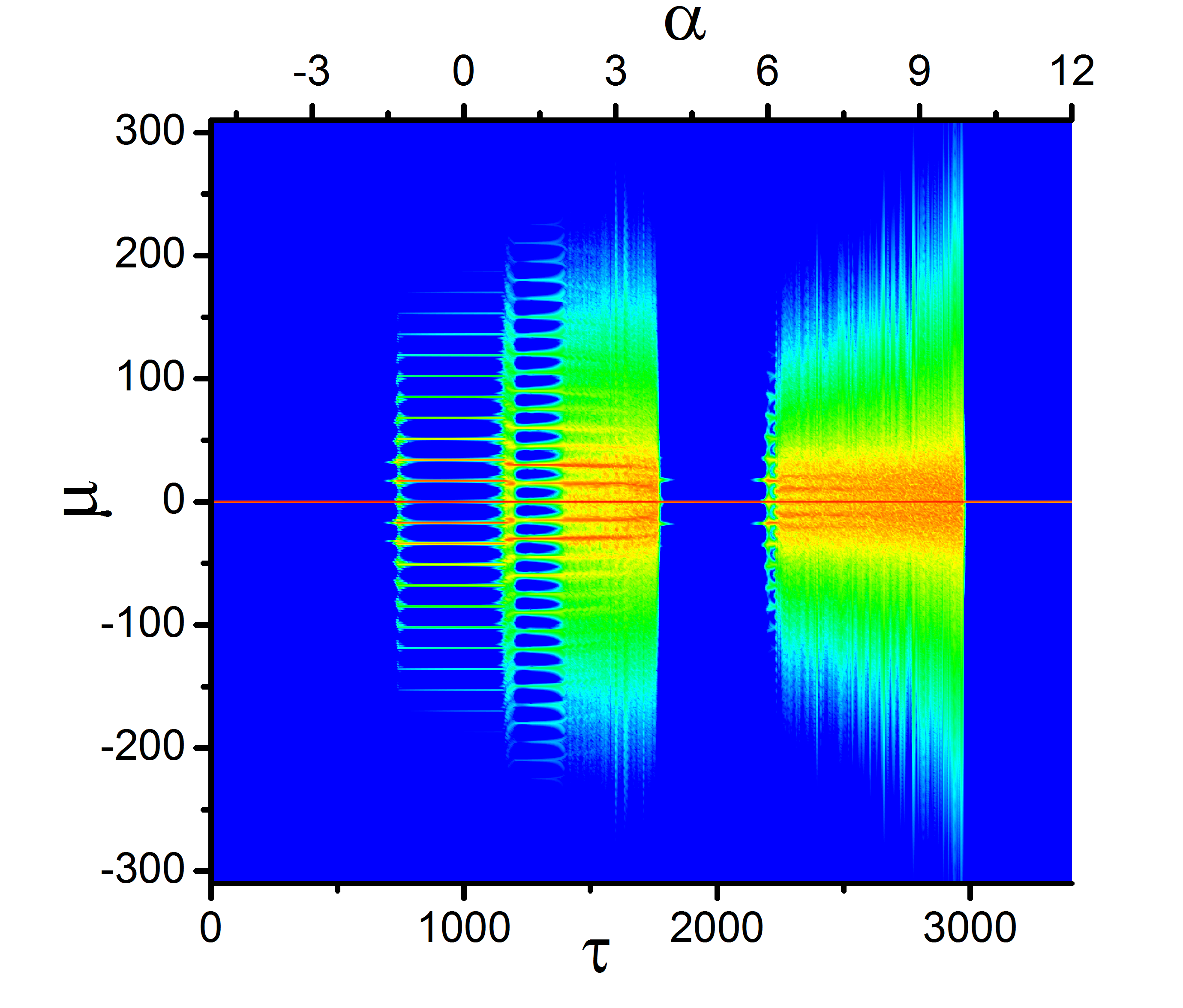}
\caption{{a)} Nonlinear double resonance at anomalous GVD: {thick} blue - stationary analytical solution (solid -- stable, dashed -- unstable, high-finesse regime), thin solid red {(dark) and yellow (light)} - solutions of \eqref{LLEBsimp} with different initial conditions.
{b)} evolution of the spectrum upon frequency scan at anomalous GVD. Plotted variables are dimensionless.}
\label{fig:dyn_anom}
\end{figure*}
\begin{figure*}[t]
\centering
\includegraphics[width=0.47\linewidth]{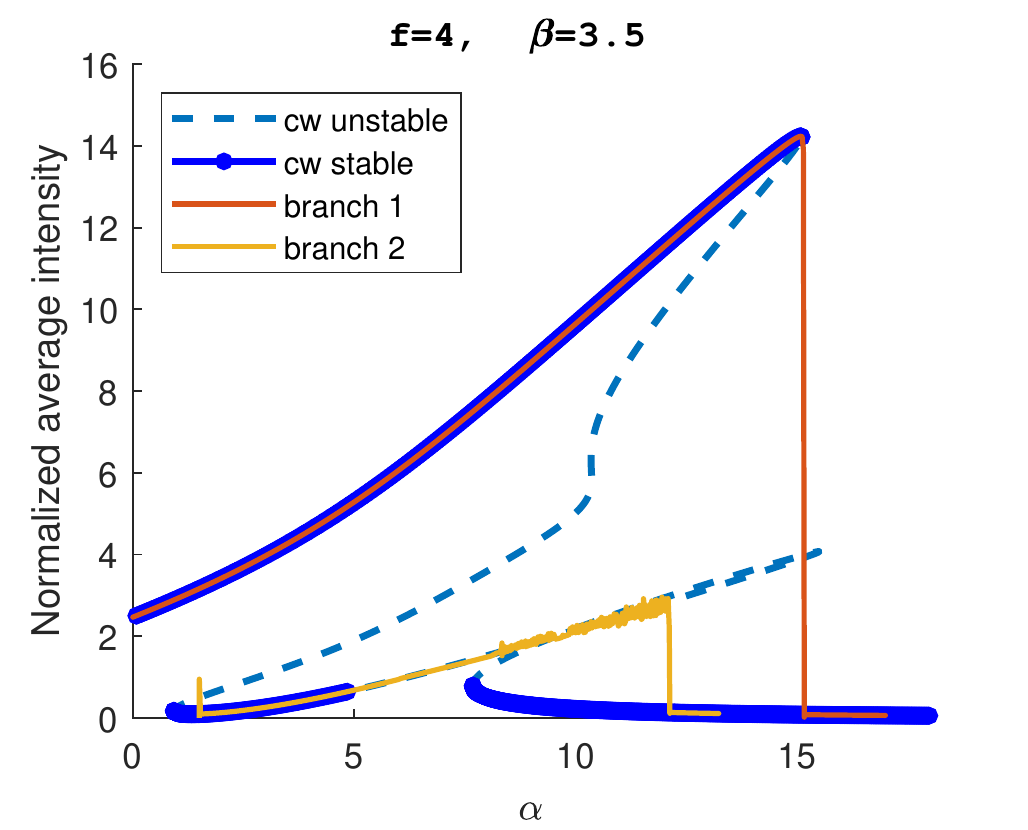}
\includegraphics[width=0.47\linewidth]{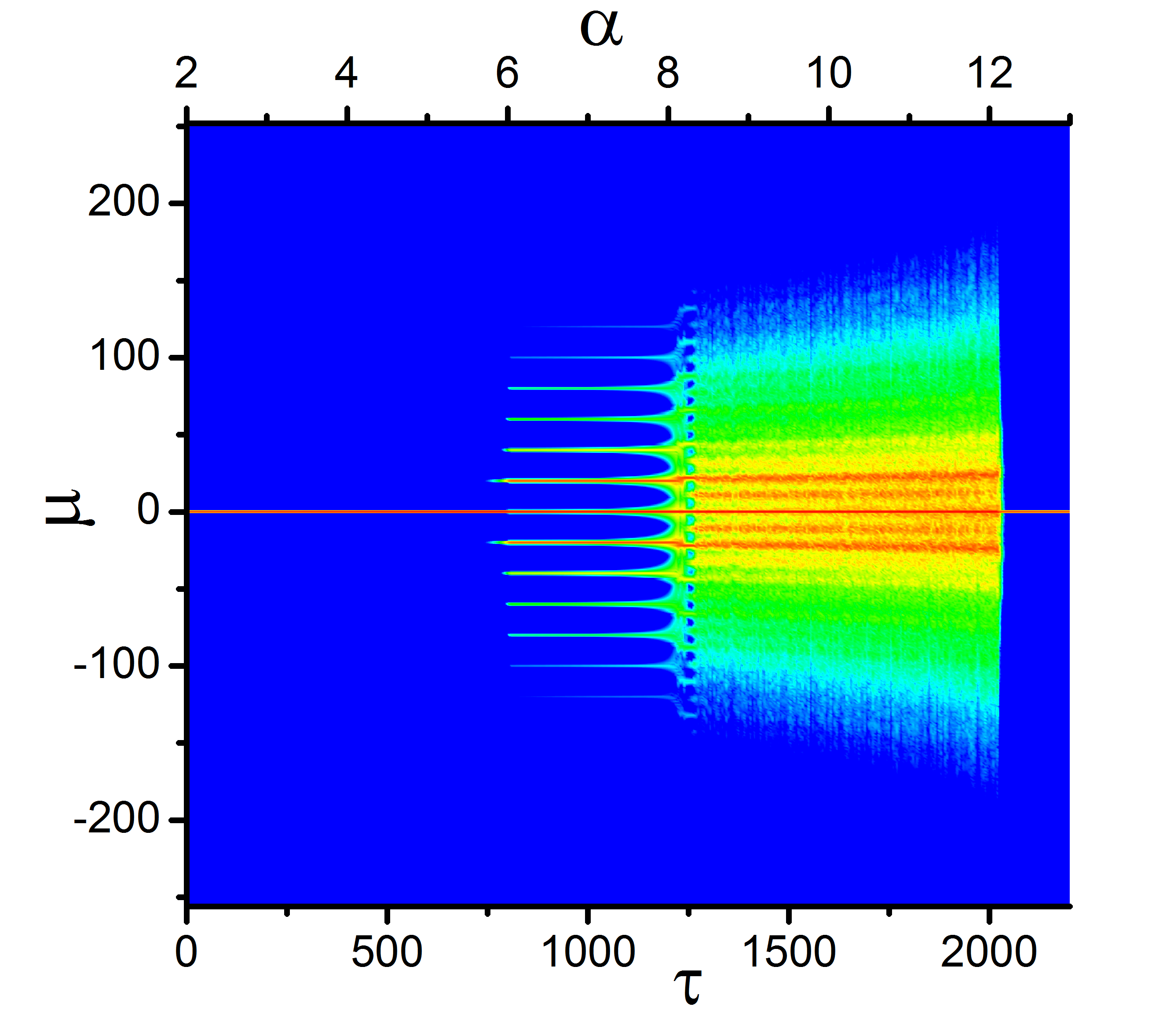}
\caption{{a)} Nonlinear double resonance at normal GVD: {thick} blue - stationary analytical solution (solid -- stable, dashed -- unstable, high-finesse regime), thin solid red {(dark) and yellow (light)} - solutions of \eqref{LLEBsimp} with different initial conditions. {b)} evolution of the spectrum upon frequency scan at normal GVD (second branch). Plotted variables are dimensionless.}
\label{fig:dyn_norm}
\end{figure*}

For considered parameters, critical finesse values are in the range 200 - 500. Before this threshold the soliton dynamics depends on $d_1$ and soliton can be unstable, form some complex patterns or experience drift (see Fig. \ref{fig:finesse}). Above critical value, soliton parameters and dynamics do not depend on the finesse value. We also found that this threshold slightly increases with $\beta$, $f$ and $\alpha$. This result is very similar to the predictions of stability analysis (see right panel of Fig. \ref{fig:charEqSimp}). Note that for typical WGM microresonator the finesse value is of the order of $10^4$ (or larger, up to the $10^7$ \cite{Savchenkov:07}) and, thus, the simplified system can be used for numerical simulations.

\begin{figure*}[ht]
\centering
\includegraphics[width=0.47\linewidth]{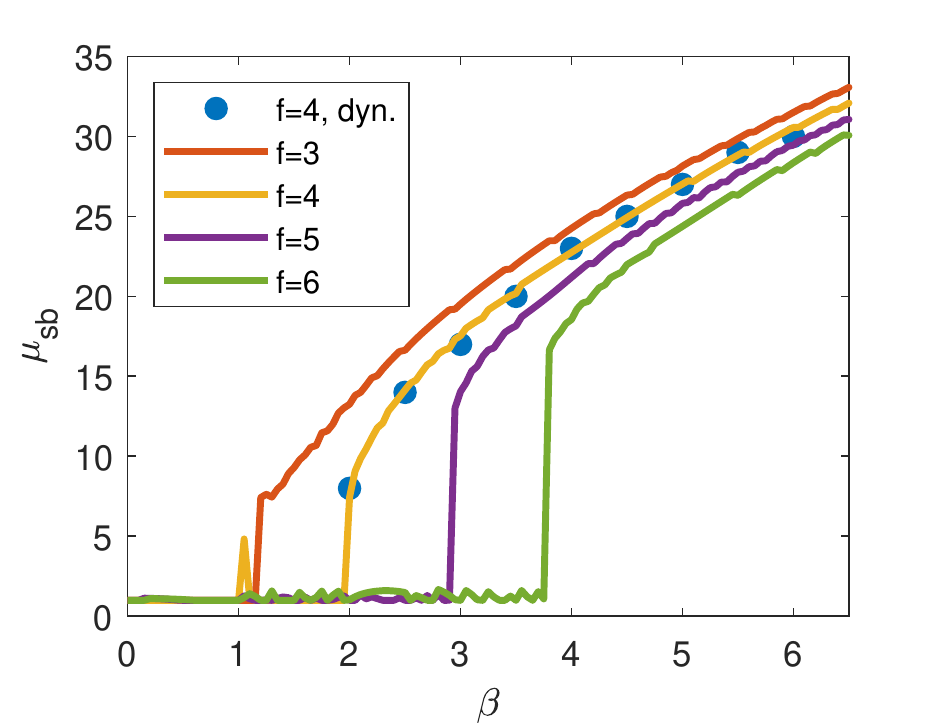}
\includegraphics[width=0.47\linewidth]{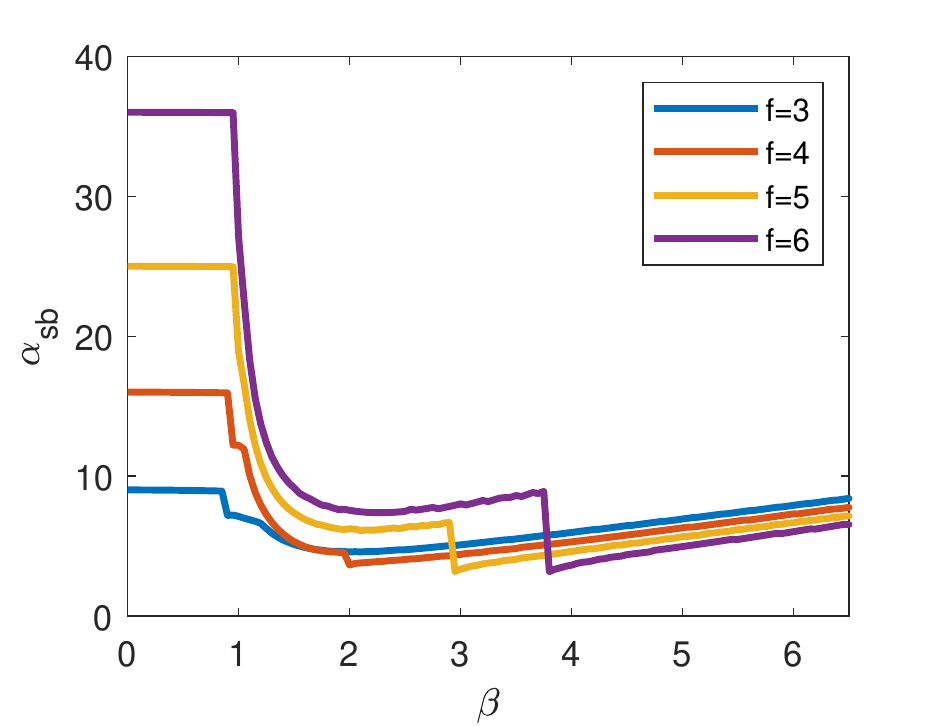}
\caption{The first sideband number $\mu_{\rm sb}$ and the detuning $\alpha_{\rm sb}$, after which it appears, vs linear coupling coefficient $\beta$. Plotted variables are dimensionless.}
\label{fig:musb}
\end{figure*}

Then we use equation \eqref{LLEBsimp} to study the dynamics of the considered system upon frequency scan ($\alpha=\alpha_0+v\tau$, $v=0.005$).
The examples of the nonlinear processes occurring during the scanning of both branches for the cases of the anomalous and normal GVD are shown in Figs. \ref{fig:dyn_anom} and \ref{fig:dyn_norm}. In the case of anomalous GVD, it is shown that when scanning the upper branch (starting with large negative detuning values), the generation of solitons is possible up to a certain critical value of the forward-backward wave coupling coefficient $\beta_{\rm dyn}$.
For $f=4$, $d_2=0.01$ this value is of the order of $\beta_{\rm dyn} \approx 1.25$. 
At values of the coupling coefficient less than this value, the influence of the backward wave is almost imperceptible. When approaching this value, a decrease in the number of generated solitons and even a single-soliton regime is observed, and when exceeded, there is no transition from the chaotic to soliton regime. Note, that similar results were demonstrated in \cite{Fujii:17}. However, accounting of the nonlinear cross-action terms, neglected in \cite{Fujii:17}, provides more accurate description of generation dynamics and more precise boundaries of soliton existence and stability domains.  With a sufficient value of the coupling coefficient, in addition to scanning the upper branch, it is possible to scan the lower branch, starting from the particular range of the detuning values (see left panel in Fig. \ref{fig:dyn_anom}). In this case, a similar nonlinear dynamics is observed on the lower branch, including the generation of primary sidebands and the chaotic regime, but the generation of solitons is absent. Moreover, if the splitting is large enough, and the scan comes from sufficiently large negative detuning values, then two frequency ranges corresponding to the upper and lower branches of the resonance curve can be observed where the frequency comb is generated (see the right panel in Fig. \ref{fig:dyn_anom}).
Note, that the search for the stationary solutions of equation \eqref{LLEBsimp} shows that solitons can exist at values $\beta_{\rm cr}>\beta_{\rm dyn}$, and this critical value $\beta_{\rm cr}$ increases with the growth of the detuning $\alpha$. For example, while critical value for the soliton excitation $\beta_{\rm dyn} \approx 1.25$, at $\alpha=12$ stable solitons exist if $\beta<2.33$ and at $\alpha=18$ - if $\beta<6.75$.  
However, these states turn out to be unattainable by the standard method of the frequency tuning. It is also interesting that, because of the integral term describing the cross-action, the existence domains for different numbers of solitons are not the same. This effect can be used for the deterministic single-soliton generation.

We also found that in the normal dispersion regime, the scanning of the main resonance does not provide generation of the additional spectral components which is consistent with results obtained from the LSA. At the same time, modulational instability is observed at the second branch of the resonance curve, which provides a new mechanism for the generation of the frequency comb (see left panel in Fig. \ref{fig:dyn_norm}). We show that this instability is connected to the loop on the tip of the second resonance which does not form for $f<2.2$ (row "loop" in Table \ref{charPoint}) and no sideband generation occurs. The second branch existence correlates with the moment when the loop becomes separated from the resonance (row "split" in the Table \ref{charPoint}).
While scanning the second branch and a certain detuning value is reached, the first sidebands appear, and then due to the non-degenerate four-wave interaction, the other frequency components are generated. Then, a chaotic regime is observed, corresponding to the generation of an incoherent comb, which then passes into a stable low-intensity single-mode state (see the right panel in Fig. \ref{fig:dyn_norm}). However, generation of solitonic pulses or platicons \cite{Lobanov:15,Lobanov2017} is not observed for  the studied parameters.
The parameters of the generated primary frequency comb (or Turing patterns in temporal representation) also depend on the linear coupling coefficient of the forward and backward waves. It was found that the distance between the pumped mode and primary sidebands increases with the linear coupling coefficient $\beta$. 
We also estimate the detuning $\alpha_{\rm sb}$ and mode number $\mu_{\rm sb}$ at which the first sideband appears during laser sweeping with the results of LSA from \eqref{charEqSimp} as the point at which $\Re[\lambda]$ becomes positive. Fig. \ref{fig:musb} shows the results together with the points, obtained by the numerical solution of \eqref{LLEBsimp}, that are in good agreement. Since $\mu$ and $d_2$ appear inside $p_m$ as a united term $d_2\mu^2$, we also find out a simple $1/\sqrt{d_2}$ scaling for $\mu_{\rm sb}$, that was also confirmed numerically.

\section{Conclusion}
Here we presented an original mathematical model describing nonlinear processes in high-Q Kerr microresonators with backscattering. Resulting system of equations was derived from the first principles and takes both linear forward to backward coupling and nonlinear cross-action into account. This new model is quite similar to the previously used ones, but has a couple of important physically justified differences, that influence the dynamics and thresholds of the nonlinear processes.
For real microresonators that usually have high finesse the system can be significantly simplified. This was checked for different combination of other parameters by means of both direct modeling and LSA. { It was shown that the bound of the high-finesse approximation is quite low, but increases with the pump power.} The nonlinear mode splitting was also analyzed and {the dependence of the resonance curve on the pump amplitude and backscattering coefficient was studied}.
Performed numerical simulations showed that the backscattering modifies the DKS existence region and at the same time provide modulational instability in the normal group velocity dispersion regime.
{Proposed model provides deep insight and accurate description of the complex dynamics of nonlinear processes in high-Q WGM microresonators and can be applied for a wide class of the spherical-symmetric WGM-like systems.}

\section{Acknowledgements}
The  authors  gratefully acknowledge help of Prof. Dmitry Skryabin for valuable discussions.

This work was supported by Russian Science Foundation (grant 17-12-01413).

\bibliography{Thebib}

\begin{thebibliography}{47}%
\makeatletter
\providecommand \@ifxundefined [1]{%
 \@ifx{#1\undefined}
}%
\providecommand \@ifnum [1]{%
 \ifnum #1\expandafter \@firstoftwo
 \else \expandafter \@secondoftwo
 \fi
}%
\providecommand \@ifx [1]{%
 \ifx #1\expandafter \@firstoftwo
 \else \expandafter \@secondoftwo
 \fi
}%
\providecommand \natexlab [1]{#1}%
\providecommand \enquote  [1]{``#1''}%
\providecommand \bibnamefont  [1]{#1}%
\providecommand \bibfnamefont [1]{#1}%
\providecommand \citenamefont [1]{#1}%
\providecommand \href@noop [0]{\@secondoftwo}%
\providecommand \href [0]{\begingroup \@sanitize@url \@href}%
\providecommand \@href[1]{\@@startlink{#1}\@@href}%
\providecommand \@@href[1]{\endgroup#1\@@endlink}%
\providecommand \@sanitize@url [0]{\catcode `\\12\catcode `\$12\catcode
  `\&12\catcode `\#12\catcode `\^12\catcode `\_12\catcode `\%12\relax}%
\providecommand \@@startlink[1]{}%
\providecommand \@@endlink[0]{}%
\providecommand \url  [0]{\begingroup\@sanitize@url \@url }%
\providecommand \@url [1]{\endgroup\@href {#1}{\urlprefix }}%
\providecommand \urlprefix  [0]{URL }%
\providecommand \Eprint [0]{\href }%
\providecommand \doibase [0]{https://doi.org/}%
\providecommand \selectlanguage [0]{\@gobble}%
\providecommand \bibinfo  [0]{\@secondoftwo}%
\providecommand \bibfield  [0]{\@secondoftwo}%
\providecommand \translation [1]{[#1]}%
\providecommand \BibitemOpen [0]{}%
\providecommand \bibitemStop [0]{}%
\providecommand \bibitemNoStop [0]{.\EOS\space}%
\providecommand \EOS [0]{\spacefactor3000\relax}%
\providecommand \BibitemShut  [1]{\csname bibitem#1\endcsname}%
\let\auto@bib@innerbib\@empty
\bibitem [{\citenamefont {{Ilchenko}}\ and\ \citenamefont
  {{Matsko}}(2006)}]{Matsko2006}%
  \BibitemOpen
  \bibfield  {author} {\bibinfo {author} {\bibfnamefont {V.~S.}\ \bibnamefont
  {{Ilchenko}}}\ and\ \bibinfo {author} {\bibfnamefont {A.~B.}\ \bibnamefont
  {{Matsko}}},\ }\bibfield  {title} {\bibinfo {title} {Optical resonators with
  whispering-gallery modes-part {II}: applications},\ }\href
  {https://doi.org/10.1109/JSTQE.2005.862943} {\bibfield  {journal} {\bibinfo
  {journal} {IEEE Journal of Selected Topics in Quantum Electronics}\ }\textbf
  {\bibinfo {volume} {12}},\ \bibinfo {pages} {15} (\bibinfo {year}
  {2006})}\BibitemShut {NoStop}%
\bibitem [{\citenamefont {Strekalov}\ \emph {et~al.}(2016)\citenamefont
  {Strekalov}, \citenamefont {Marquardt}, \citenamefont {Matsko}, \citenamefont
  {Schwefel},\ and\ \citenamefont {Leuchs}}]{Strekalov_2016}%
  \BibitemOpen
  \bibfield  {author} {\bibinfo {author} {\bibfnamefont {D.~V.}\ \bibnamefont
  {Strekalov}}, \bibinfo {author} {\bibfnamefont {C.}~\bibnamefont
  {Marquardt}}, \bibinfo {author} {\bibfnamefont {A.~B.}\ \bibnamefont
  {Matsko}}, \bibinfo {author} {\bibfnamefont {H.~G.~L.}\ \bibnamefont
  {Schwefel}},\ and\ \bibinfo {author} {\bibfnamefont {G.}~\bibnamefont
  {Leuchs}},\ }\bibfield  {title} {\bibinfo {title} {Nonlinear and quantum
  optics with whispering gallery resonators},\ }\href
  {https://doi.org/10.1088/2040-8978/18/12/123002} {\bibfield  {journal}
  {\bibinfo  {journal} {Journal of Optics}\ }\textbf {\bibinfo {volume} {18}},\
  \bibinfo {pages} {123002} (\bibinfo {year} {2016})}\BibitemShut {NoStop}%
\bibitem [{\citenamefont {Lin}\ \emph {et~al.}(2017)\citenamefont {Lin},
  \citenamefont {Coillet},\ and\ \citenamefont {Chembo}}]{Lin:17}%
  \BibitemOpen
  \bibfield  {author} {\bibinfo {author} {\bibfnamefont {G.}~\bibnamefont
  {Lin}}, \bibinfo {author} {\bibfnamefont {A.}~\bibnamefont {Coillet}},\ and\
  \bibinfo {author} {\bibfnamefont {Y.~K.}\ \bibnamefont {Chembo}},\ }\bibfield
   {title} {\bibinfo {title} {Nonlinear photonics with high-{Q}
  whispering-gallery-mode resonators},\ }\href
  {https://doi.org/10.1364/AOP.9.000828} {\bibfield  {journal} {\bibinfo
  {journal} {Adv. Opt. Photon.}\ }\textbf {\bibinfo {volume} {9}},\ \bibinfo
  {pages} {828} (\bibinfo {year} {2017})}\BibitemShut {NoStop}%
\bibitem [{\citenamefont {Del’Haye}\ \emph {et~al.}(2007)\citenamefont
  {Del’Haye}, \citenamefont {Schliesser}, \citenamefont {Arcizet},
  \citenamefont {Wilken}, \citenamefont {Holzwarth},\ and\ \citenamefont
  {Kippenberg}}]{Kippenberg-2007}%
  \BibitemOpen
  \bibfield  {author} {\bibinfo {author} {\bibfnamefont {P.}~\bibnamefont
  {Del’Haye}}, \bibinfo {author} {\bibfnamefont {A.}~\bibnamefont
  {Schliesser}}, \bibinfo {author} {\bibfnamefont {O.}~\bibnamefont {Arcizet}},
  \bibinfo {author} {\bibfnamefont {T.}~\bibnamefont {Wilken}}, \bibinfo
  {author} {\bibfnamefont {R.}~\bibnamefont {Holzwarth}},\ and\ \bibinfo
  {author} {\bibfnamefont {T.~J.}\ \bibnamefont {Kippenberg}},\ }\bibfield
  {title} {\bibinfo {title} {Optical frequency comb generation from a
  monolithic microresonator},\ }\href {https://doi.org/10.1038/nature06401}
  {\bibfield  {journal} {\bibinfo  {journal} {Nature}\ }\textbf {\bibinfo
  {volume} {450}},\ \bibinfo {pages} {1214} (\bibinfo {year}
  {2007})}\BibitemShut {NoStop}%
\bibitem [{\citenamefont {Pasquazi}\ \emph {et~al.}(2018)\citenamefont
  {Pasquazi}, \citenamefont {Peccianti}, \citenamefont {Razzari}, \citenamefont
  {Moss}, \citenamefont {Coen}, \citenamefont {Erkintalo}, \citenamefont
  {Chembo}, \citenamefont {Hansson}, \citenamefont {Wabnitz}, \citenamefont
  {Del’Haye}, \citenamefont {Xue}, \citenamefont {Weiner},\ and\
  \citenamefont {Morandotti}}]{PASQUAZI20181}%
  \BibitemOpen
  \bibfield  {author} {\bibinfo {author} {\bibfnamefont {A.}~\bibnamefont
  {Pasquazi}}, \bibinfo {author} {\bibfnamefont {M.}~\bibnamefont {Peccianti}},
  \bibinfo {author} {\bibfnamefont {L.}~\bibnamefont {Razzari}}, \bibinfo
  {author} {\bibfnamefont {D.~J.}\ \bibnamefont {Moss}}, \bibinfo {author}
  {\bibfnamefont {S.}~\bibnamefont {Coen}}, \bibinfo {author} {\bibfnamefont
  {M.}~\bibnamefont {Erkintalo}}, \bibinfo {author} {\bibfnamefont {Y.~K.}\
  \bibnamefont {Chembo}}, \bibinfo {author} {\bibfnamefont {T.}~\bibnamefont
  {Hansson}}, \bibinfo {author} {\bibfnamefont {S.}~\bibnamefont {Wabnitz}},
  \bibinfo {author} {\bibfnamefont {P.}~\bibnamefont {Del’Haye}}, \bibinfo
  {author} {\bibfnamefont {X.}~\bibnamefont {Xue}}, \bibinfo {author}
  {\bibfnamefont {A.~M.}\ \bibnamefont {Weiner}},\ and\ \bibinfo {author}
  {\bibfnamefont {R.}~\bibnamefont {Morandotti}},\ }\bibfield  {title}
  {\bibinfo {title} {Micro-combs: A novel generation of optical sources},\
  }\href {https://doi.org/https://doi.org/10.1016/j.physrep.2017.08.004}
  {\bibfield  {journal} {\bibinfo  {journal} {Physics Reports}\ }\textbf
  {\bibinfo {volume} {729}},\ \bibinfo {pages} {1 } (\bibinfo {year}
  {2018})}\BibitemShut {NoStop}%
\bibitem [{\citenamefont {Gaeta}\ \emph {et~al.}(2019)\citenamefont {Gaeta},
  \citenamefont {Lipson},\ and\ \citenamefont {Kippenberg}}]{Gaeta2019}%
  \BibitemOpen
  \bibfield  {author} {\bibinfo {author} {\bibfnamefont {A.}~\bibnamefont
  {Gaeta}}, \bibinfo {author} {\bibfnamefont {M.}~\bibnamefont {Lipson}},\ and\
  \bibinfo {author} {\bibfnamefont {T.}~\bibnamefont {Kippenberg}},\ }\bibfield
   {title} {\bibinfo {title} {Photonic-chip-based frequency combs},\ }\href
  {https://doi.org/10.1038/s41566-019-0358-x} {\bibfield  {journal} {\bibinfo
  {journal} {Nature Photon.}\ }\textbf {\bibinfo {volume} {13}},\ \bibinfo
  {pages} {158–169} (\bibinfo {year} {2019})}\BibitemShut {NoStop}%
\bibitem [{\citenamefont {Herr}\ \emph {et~al.}(2014)\citenamefont {Herr},
  \citenamefont {Brasch}, \citenamefont {Jost}, \citenamefont {Wang},
  \citenamefont {Kondratiev}, \citenamefont {Gorodetsky},\ and\ \citenamefont
  {Kippenberg}}]{herr2014temporal}%
  \BibitemOpen
  \bibfield  {author} {\bibinfo {author} {\bibfnamefont {T.}~\bibnamefont
  {Herr}}, \bibinfo {author} {\bibfnamefont {V.}~\bibnamefont {Brasch}},
  \bibinfo {author} {\bibfnamefont {J.~D.}\ \bibnamefont {Jost}}, \bibinfo
  {author} {\bibfnamefont {C.~Y.}\ \bibnamefont {Wang}}, \bibinfo {author}
  {\bibfnamefont {N.~M.}\ \bibnamefont {Kondratiev}}, \bibinfo {author}
  {\bibfnamefont {M.~L.}\ \bibnamefont {Gorodetsky}},\ and\ \bibinfo {author}
  {\bibfnamefont {T.~J.}\ \bibnamefont {Kippenberg}},\ }\bibfield  {title}
  {\bibinfo {title} {Temporal solitons in optical microresonators},\ }\href
  {https://doi.org/10.1038/nphoton.2013.343} {\bibfield  {journal} {\bibinfo
  {journal} {Nat. Photon.}\ }\textbf {\bibinfo {volume} {8}},\ \bibinfo {pages}
  {145} (\bibinfo {year} {2014})}\BibitemShut {NoStop}%
\bibitem [{\citenamefont {Kippenberg}\ \emph {et~al.}(2018)\citenamefont
  {Kippenberg}, \citenamefont {Gaeta}, \citenamefont {Lipson},\ and\
  \citenamefont {Gorodetsky}}]{Kippenbergeaan8083}%
  \BibitemOpen
  \bibfield  {author} {\bibinfo {author} {\bibfnamefont {T.~J.}\ \bibnamefont
  {Kippenberg}}, \bibinfo {author} {\bibfnamefont {A.~L.}\ \bibnamefont
  {Gaeta}}, \bibinfo {author} {\bibfnamefont {M.}~\bibnamefont {Lipson}},\ and\
  \bibinfo {author} {\bibfnamefont {M.~L.}\ \bibnamefont {Gorodetsky}},\
  }\bibfield  {title} {\bibinfo {title} {Dissipative {K}err solitons in optical
  microresonators},\ }\href {https://doi.org/10.1126/science.aan8083}
  {\bibfield  {journal} {\bibinfo  {journal} {Science}\ }\textbf {\bibinfo
  {volume} {361}},\ \bibinfo {pages} {eaan8083} (\bibinfo {year}
  {2018})}\BibitemShut {NoStop}%
\bibitem [{\citenamefont {Vassiliev}\ \emph {et~al.}(1998)\citenamefont
  {Vassiliev}, \citenamefont {Velichansky}, \citenamefont {Ilchenko},
  \citenamefont {Gorodetsky}, \citenamefont {Hollberg},\ and\ \citenamefont
  {Yarovitsky}}]{VASSILIEV1998305}%
  \BibitemOpen
  \bibfield  {author} {\bibinfo {author} {\bibfnamefont {V.}~\bibnamefont
  {Vassiliev}}, \bibinfo {author} {\bibfnamefont {V.}~\bibnamefont
  {Velichansky}}, \bibinfo {author} {\bibfnamefont {V.}~\bibnamefont
  {Ilchenko}}, \bibinfo {author} {\bibfnamefont {M.}~\bibnamefont
  {Gorodetsky}}, \bibinfo {author} {\bibfnamefont {L.}~\bibnamefont
  {Hollberg}},\ and\ \bibinfo {author} {\bibfnamefont {A.}~\bibnamefont
  {Yarovitsky}},\ }\bibfield  {title} {\bibinfo {title} {Narrow-line-width
  diode laser with a high-{Q} microsphere resonator},\ }\href
  {https://doi.org/https://doi.org/10.1016/S0030-4018(98)00578-1} {\bibfield
  {journal} {\bibinfo  {journal} {Optics Communications}\ }\textbf {\bibinfo
  {volume} {158}},\ \bibinfo {pages} {305 } (\bibinfo {year}
  {1998})}\BibitemShut {NoStop}%
\bibitem [{\citenamefont {Kondratiev}\ \emph {et~al.}(2017)\citenamefont
  {Kondratiev}, \citenamefont {Lobanov}, \citenamefont {Cherenkov},
  \citenamefont {Voloshin}, \citenamefont {Pavlov}, \citenamefont {Koptyaev},\
  and\ \citenamefont {Gorodetsky}}]{Kondratiev:17}%
  \BibitemOpen
  \bibfield  {author} {\bibinfo {author} {\bibfnamefont {N.~M.}\ \bibnamefont
  {Kondratiev}}, \bibinfo {author} {\bibfnamefont {V.~E.}\ \bibnamefont
  {Lobanov}}, \bibinfo {author} {\bibfnamefont {A.~V.}\ \bibnamefont
  {Cherenkov}}, \bibinfo {author} {\bibfnamefont {A.~S.}\ \bibnamefont
  {Voloshin}}, \bibinfo {author} {\bibfnamefont {N.~G.}\ \bibnamefont
  {Pavlov}}, \bibinfo {author} {\bibfnamefont {S.}~\bibnamefont {Koptyaev}},\
  and\ \bibinfo {author} {\bibfnamefont {M.~L.}\ \bibnamefont {Gorodetsky}},\
  }\bibfield  {title} {\bibinfo {title} {Self-injection locking of a laser
  diode to a high-{Q} {WGM} microresonator},\ }\href
  {https://doi.org/10.1364/OE.25.028167} {\bibfield  {journal} {\bibinfo
  {journal} {Opt. Express}\ }\textbf {\bibinfo {volume} {25}},\ \bibinfo
  {pages} {28167} (\bibinfo {year} {2017})}\BibitemShut {NoStop}%
\bibitem [{\citenamefont {Galiev}\ \emph {et~al.}(2018)\citenamefont {Galiev},
  \citenamefont {Pavlov}, \citenamefont {Kondratiev}, \citenamefont {Koptyaev},
  \citenamefont {Lobanov}, \citenamefont {Voloshin}, \citenamefont
  {Gorodnitskiy},\ and\ \citenamefont {Gorodetsky}}]{Galiev:18}%
  \BibitemOpen
  \bibfield  {author} {\bibinfo {author} {\bibfnamefont {R.~R.}\ \bibnamefont
  {Galiev}}, \bibinfo {author} {\bibfnamefont {N.~G.}\ \bibnamefont {Pavlov}},
  \bibinfo {author} {\bibfnamefont {N.~M.}\ \bibnamefont {Kondratiev}},
  \bibinfo {author} {\bibfnamefont {S.}~\bibnamefont {Koptyaev}}, \bibinfo
  {author} {\bibfnamefont {V.~E.}\ \bibnamefont {Lobanov}}, \bibinfo {author}
  {\bibfnamefont {A.~S.}\ \bibnamefont {Voloshin}}, \bibinfo {author}
  {\bibfnamefont {A.~S.}\ \bibnamefont {Gorodnitskiy}},\ and\ \bibinfo {author}
  {\bibfnamefont {M.~L.}\ \bibnamefont {Gorodetsky}},\ }\bibfield  {title}
  {\bibinfo {title} {Spectrum collapse, narrow linewidth, and bogatov effect in
  diode lasers locked to high-{Q} optical microresonators},\ }\href
  {https://doi.org/10.1364/OE.26.030509} {\bibfield  {journal} {\bibinfo
  {journal} {Opt. Express}\ }\textbf {\bibinfo {volume} {26}},\ \bibinfo
  {pages} {30509} (\bibinfo {year} {2018})}\BibitemShut {NoStop}%
\bibitem [{\citenamefont {Savchenkov}\ \emph {et~al.}(2018)\citenamefont
  {Savchenkov}, \citenamefont {Williams},\ and\ \citenamefont
  {Matsko}}]{photonics5040043}%
  \BibitemOpen
  \bibfield  {author} {\bibinfo {author} {\bibfnamefont {A.}~\bibnamefont
  {Savchenkov}}, \bibinfo {author} {\bibfnamefont {S.}~\bibnamefont
  {Williams}},\ and\ \bibinfo {author} {\bibfnamefont {A.}~\bibnamefont
  {Matsko}},\ }\bibfield  {title} {\bibinfo {title} {On stiffness of optical
  self-injection locking},\ }\bibfield  {journal} {\bibinfo  {journal}
  {Photonics}\ }\textbf {\bibinfo {volume} {5}},\ \href
  {https://doi.org/10.3390/photonics5040043} {10.3390/photonics5040043}
  (\bibinfo {year} {2018})\BibitemShut {NoStop}%
\bibitem [{\citenamefont {Sprenger}\ \emph {et~al.}(2009)\citenamefont
  {Sprenger}, \citenamefont {Schwefel},\ and\ \citenamefont
  {Wang}}]{Sprenger:09}%
  \BibitemOpen
  \bibfield  {author} {\bibinfo {author} {\bibfnamefont {B.}~\bibnamefont
  {Sprenger}}, \bibinfo {author} {\bibfnamefont {H.~G.~L.}\ \bibnamefont
  {Schwefel}},\ and\ \bibinfo {author} {\bibfnamefont {L.~J.}\ \bibnamefont
  {Wang}},\ }\bibfield  {title} {\bibinfo {title}
  {Whispering-gallery-mode-resonator-stabilized narrow-linewidth fiber loop
  laser},\ }\href {https://doi.org/10.1364/OL.34.003370} {\bibfield  {journal}
  {\bibinfo  {journal} {Opt. Lett.}\ }\textbf {\bibinfo {volume} {34}},\
  \bibinfo {pages} {3370} (\bibinfo {year} {2009})}\BibitemShut {NoStop}%
\bibitem [{\citenamefont {Liang}\ \emph {et~al.}(2010)\citenamefont {Liang},
  \citenamefont {Ilchenko}, \citenamefont {Savchenkov}, \citenamefont {Matsko},
  \citenamefont {Seidel},\ and\ \citenamefont {Maleki}}]{Liang:10}%
  \BibitemOpen
  \bibfield  {author} {\bibinfo {author} {\bibfnamefont {W.}~\bibnamefont
  {Liang}}, \bibinfo {author} {\bibfnamefont {V.~S.}\ \bibnamefont {Ilchenko}},
  \bibinfo {author} {\bibfnamefont {A.~A.}\ \bibnamefont {Savchenkov}},
  \bibinfo {author} {\bibfnamefont {A.~B.}\ \bibnamefont {Matsko}}, \bibinfo
  {author} {\bibfnamefont {D.}~\bibnamefont {Seidel}},\ and\ \bibinfo {author}
  {\bibfnamefont {L.}~\bibnamefont {Maleki}},\ }\bibfield  {title} {\bibinfo
  {title} {Whispering-gallery-mode-resonator-based ultranarrow linewidth
  external-cavity semiconductor laser},\ }\href
  {https://doi.org/10.1364/OL.35.002822} {\bibfield  {journal} {\bibinfo
  {journal} {Opt. Lett.}\ }\textbf {\bibinfo {volume} {35}},\ \bibinfo {pages}
  {2822} (\bibinfo {year} {2010})}\BibitemShut {NoStop}%
\bibitem [{\citenamefont {Pavlov}\ \emph {et~al.}(2018)\citenamefont {Pavlov},
  \citenamefont {Koptyaev}, \citenamefont {Lihachev}, \citenamefont {Voloshin},
  \citenamefont {Gorodnitskiy}, \citenamefont {Ryabko}, \citenamefont
  {Polonsky},\ and\ \citenamefont {Gorodetsky}}]{Pavlov_18np}%
  \BibitemOpen
  \bibfield  {author} {\bibinfo {author} {\bibfnamefont {N.~G.}\ \bibnamefont
  {Pavlov}}, \bibinfo {author} {\bibfnamefont {S.}~\bibnamefont {Koptyaev}},
  \bibinfo {author} {\bibfnamefont {G.~V.}\ \bibnamefont {Lihachev}}, \bibinfo
  {author} {\bibfnamefont {A.~S.}\ \bibnamefont {Voloshin}}, \bibinfo {author}
  {\bibfnamefont {A.~A.}\ \bibnamefont {Gorodnitskiy}}, \bibinfo {author}
  {\bibfnamefont {M.~V.}\ \bibnamefont {Ryabko}}, \bibinfo {author}
  {\bibfnamefont {S.~V.}\ \bibnamefont {Polonsky}},\ and\ \bibinfo {author}
  {\bibfnamefont {M.~L.}\ \bibnamefont {Gorodetsky}},\ }\bibfield  {title}
  {\bibinfo {title} {Narrow linewidth lasing and soliton {K}err-microcombs with
  ordinary laser diodes},\ }\href {https://doi.org/10.1038/s41566-018-0277-2}
  {\bibfield  {journal} {\bibinfo  {journal} {Nat. Photon.}\ }\textbf {\bibinfo
  {volume} {12}},\ \bibinfo {pages} {694} (\bibinfo {year} {2018})}\BibitemShut
  {NoStop}%
\bibitem [{\citenamefont {Raja}\ \emph {et~al.}(2019)\citenamefont {Raja},
  \citenamefont {Voloshin}, \citenamefont {Guo}, \citenamefont {Agafonova},
  \citenamefont {Liu}, \citenamefont {Gorodnitskiy}, \citenamefont {Karpov},
  \citenamefont {Pavlov}, \citenamefont {Lucas}, \citenamefont {Galiev},
  \citenamefont {Shitikov}, \citenamefont {Jost}, \citenamefont {Gorodetsky},\
  and\ \citenamefont {Kippenberg}}]{Raja2019}%
  \BibitemOpen
  \bibfield  {author} {\bibinfo {author} {\bibfnamefont {A.~S.}\ \bibnamefont
  {Raja}}, \bibinfo {author} {\bibfnamefont {A.~S.}\ \bibnamefont {Voloshin}},
  \bibinfo {author} {\bibfnamefont {H.}~\bibnamefont {Guo}}, \bibinfo {author}
  {\bibfnamefont {S.~E.}\ \bibnamefont {Agafonova}}, \bibinfo {author}
  {\bibfnamefont {J.}~\bibnamefont {Liu}}, \bibinfo {author} {\bibfnamefont
  {A.~S.}\ \bibnamefont {Gorodnitskiy}}, \bibinfo {author} {\bibfnamefont
  {M.}~\bibnamefont {Karpov}}, \bibinfo {author} {\bibfnamefont {N.~G.}\
  \bibnamefont {Pavlov}}, \bibinfo {author} {\bibfnamefont {E.}~\bibnamefont
  {Lucas}}, \bibinfo {author} {\bibfnamefont {R.~R.}\ \bibnamefont {Galiev}},
  \bibinfo {author} {\bibfnamefont {A.~E.}\ \bibnamefont {Shitikov}}, \bibinfo
  {author} {\bibfnamefont {J.~D.}\ \bibnamefont {Jost}}, \bibinfo {author}
  {\bibfnamefont {M.~L.}\ \bibnamefont {Gorodetsky}},\ and\ \bibinfo {author}
  {\bibfnamefont {T.~J.}\ \bibnamefont {Kippenberg}},\ }\bibfield  {title}
  {\bibinfo {title} {Electrically pumped photonic integrated soliton
  microcomb},\ }\href {https://doi.org/10.1038/s41467-019-08498-2} {\bibfield
  {journal} {\bibinfo  {journal} {Nature Communications}\ }\textbf {\bibinfo
  {volume} {10}},\ \bibinfo {pages} {690} (\bibinfo {year} {2019})}\BibitemShut
  {NoStop}%
\bibitem [{\citenamefont {Gorodetsky}\ \emph {et~al.}(2000)\citenamefont
  {Gorodetsky}, \citenamefont {Pryamikov},\ and\ \citenamefont
  {Ilchenko}}]{Gorodetsky:00}%
  \BibitemOpen
  \bibfield  {author} {\bibinfo {author} {\bibfnamefont {M.~L.}\ \bibnamefont
  {Gorodetsky}}, \bibinfo {author} {\bibfnamefont {A.~D.}\ \bibnamefont
  {Pryamikov}},\ and\ \bibinfo {author} {\bibfnamefont {V.~S.}\ \bibnamefont
  {Ilchenko}},\ }\bibfield  {title} {\bibinfo {title} {Rayleigh scattering in
  high-{Q} microspheres},\ }\href {https://doi.org/10.1364/JOSAB.17.001051}
  {\bibfield  {journal} {\bibinfo  {journal} {J. Opt. Soc. Am. B}\ }\textbf
  {\bibinfo {volume} {17}},\ \bibinfo {pages} {1051} (\bibinfo {year}
  {2000})}\BibitemShut {NoStop}%
\bibitem [{\citenamefont {Agrawal}(1987)}]{Agrawal1987}%
  \BibitemOpen
  \bibfield  {author} {\bibinfo {author} {\bibfnamefont {G.}~\bibnamefont
  {Agrawal}},\ }\bibfield  {title} {\bibinfo {title} {Modulation instability
  induced by cross-phase modulation},\ }\href
  {https://doi.org/10.1103/PhysRevLett.59.880} {\bibfield  {journal} {\bibinfo
  {journal} {Phys. Rev. Lett.}\ }\textbf {\bibinfo {volume} {59}},\ \bibinfo
  {pages} {880} (\bibinfo {year} {1987})}\BibitemShut {NoStop}%
\bibitem [{\citenamefont {Zhang}\ \emph {et~al.}(2005)\citenamefont {Zhang},
  \citenamefont {Lu}, \citenamefont {Xu},\ and\ \citenamefont
  {Wang}}]{ZHANG2005193}%
  \BibitemOpen
  \bibfield  {author} {\bibinfo {author} {\bibfnamefont {S.}~\bibnamefont
  {Zhang}}, \bibinfo {author} {\bibfnamefont {F.}~\bibnamefont {Lu}}, \bibinfo
  {author} {\bibfnamefont {W.}~\bibnamefont {Xu}},\ and\ \bibinfo {author}
  {\bibfnamefont {J.}~\bibnamefont {Wang}},\ }\bibfield  {title} {\bibinfo
  {title} {Modulation instability induced by cross-phase modulation in
  decreasing dispersion fiber},\ }\href
  {https://doi.org/https://doi.org/10.1016/j.yofte.2004.09.008} {\bibfield
  {journal} {\bibinfo  {journal} {Optical Fiber Technology}\ }\textbf {\bibinfo
  {volume} {11}},\ \bibinfo {pages} {193 } (\bibinfo {year}
  {2005})}\BibitemShut {NoStop}%
\bibitem [{\citenamefont {Tanemura}\ and\ \citenamefont
  {Kikuchi}(2003)}]{Tanemura:03}%
  \BibitemOpen
  \bibfield  {author} {\bibinfo {author} {\bibfnamefont {T.}~\bibnamefont
  {Tanemura}}\ and\ \bibinfo {author} {\bibfnamefont {K.}~\bibnamefont
  {Kikuchi}},\ }\bibfield  {title} {\bibinfo {title} {Unified analysis of
  modulational instability induced by cross-phase modulation in optical
  fibers},\ }\href {https://doi.org/10.1364/JOSAB.20.002502} {\bibfield
  {journal} {\bibinfo  {journal} {J. Opt. Soc. Am. B}\ }\textbf {\bibinfo
  {volume} {20}},\ \bibinfo {pages} {2502} (\bibinfo {year}
  {2003})}\BibitemShut {NoStop}%
\bibitem [{\citenamefont {Li}\ \emph {et~al.}(2019)\citenamefont {Li},
  \citenamefont {Leng}, \citenamefont {Zhou},\ and\ \citenamefont
  {Chen}}]{Li:19}%
  \BibitemOpen
  \bibfield  {author} {\bibinfo {author} {\bibfnamefont {L.}~\bibnamefont
  {Li}}, \bibinfo {author} {\bibfnamefont {J.}~\bibnamefont {Leng}}, \bibinfo
  {author} {\bibfnamefont {P.}~\bibnamefont {Zhou}},\ and\ \bibinfo {author}
  {\bibfnamefont {J.}~\bibnamefont {Chen}},\ }\bibfield  {title} {\bibinfo
  {title} {Modulation instability induced by intermodal cross-phase modulation
  in step-index multimode fiber},\ }\href
  {https://doi.org/10.1364/AO.58.004283} {\bibfield  {journal} {\bibinfo
  {journal} {Appl. Opt.}\ }\textbf {\bibinfo {volume} {58}},\ \bibinfo {pages}
  {4283} (\bibinfo {year} {2019})}\BibitemShut {NoStop}%
\bibitem [{\citenamefont {Jang}\ \emph {et~al.}(2016)\citenamefont {Jang},
  \citenamefont {Okawachi}, \citenamefont {Yu}, \citenamefont {Luke},
  \citenamefont {Ji}, \citenamefont {Lipson},\ and\ \citenamefont
  {Gaeta}}]{Jang:16}%
  \BibitemOpen
  \bibfield  {author} {\bibinfo {author} {\bibfnamefont {J.~K.}\ \bibnamefont
  {Jang}}, \bibinfo {author} {\bibfnamefont {Y.}~\bibnamefont {Okawachi}},
  \bibinfo {author} {\bibfnamefont {M.}~\bibnamefont {Yu}}, \bibinfo {author}
  {\bibfnamefont {K.}~\bibnamefont {Luke}}, \bibinfo {author} {\bibfnamefont
  {X.}~\bibnamefont {Ji}}, \bibinfo {author} {\bibfnamefont {M.}~\bibnamefont
  {Lipson}},\ and\ \bibinfo {author} {\bibfnamefont {A.~L.}\ \bibnamefont
  {Gaeta}},\ }\bibfield  {title} {\bibinfo {title} {Dynamics of
  mode-coupling-induced microresonator frequency combs in normal dispersion},\
  }\href {https://doi.org/10.1364/OE.24.028794} {\bibfield  {journal} {\bibinfo
   {journal} {Opt. Express}\ }\textbf {\bibinfo {volume} {24}},\ \bibinfo
  {pages} {28794} (\bibinfo {year} {2016})}\BibitemShut {NoStop}%
\bibitem [{\citenamefont {Xue}\ \emph {et~al.}(2015)\citenamefont {Xue},
  \citenamefont {Xuan}, \citenamefont {Wang}, \citenamefont {Liu},
  \citenamefont {Leaird}, \citenamefont {Qi},\ and\ \citenamefont
  {Weiner}}]{Xue-2015}%
  \BibitemOpen
  \bibfield  {author} {\bibinfo {author} {\bibfnamefont {X.}~\bibnamefont
  {Xue}}, \bibinfo {author} {\bibfnamefont {Y.}~\bibnamefont {Xuan}}, \bibinfo
  {author} {\bibfnamefont {P.-H.}\ \bibnamefont {Wang}}, \bibinfo {author}
  {\bibfnamefont {Y.}~\bibnamefont {Liu}}, \bibinfo {author} {\bibfnamefont
  {D.~E.}\ \bibnamefont {Leaird}}, \bibinfo {author} {\bibfnamefont
  {M.}~\bibnamefont {Qi}},\ and\ \bibinfo {author} {\bibfnamefont {A.~M.}\
  \bibnamefont {Weiner}},\ }\bibfield  {title} {\bibinfo {title}
  {Normal-dispersion microcombs enabled by controllable mode interactions},\
  }\href {https://doi.org/10.1002/lpor.201500107} {\bibfield  {journal}
  {\bibinfo  {journal} {Laser \& Photonics Reviews}\ }\textbf {\bibinfo
  {volume} {9}},\ \bibinfo {pages} {L23} (\bibinfo {year} {2015})}\BibitemShut
  {NoStop}%
\bibitem [{\citenamefont {Ramelow}\ \emph {et~al.}(2014)\citenamefont
  {Ramelow}, \citenamefont {Farsi}, \citenamefont {Clemmen}, \citenamefont
  {Levy}, \citenamefont {Johnson}, \citenamefont {Okawachi}, \citenamefont
  {Lamont}, \citenamefont {Lipson},\ and\ \citenamefont {Gaeta}}]{Ramelow:14}%
  \BibitemOpen
  \bibfield  {author} {\bibinfo {author} {\bibfnamefont {S.}~\bibnamefont
  {Ramelow}}, \bibinfo {author} {\bibfnamefont {A.}~\bibnamefont {Farsi}},
  \bibinfo {author} {\bibfnamefont {S.}~\bibnamefont {Clemmen}}, \bibinfo
  {author} {\bibfnamefont {J.~S.}\ \bibnamefont {Levy}}, \bibinfo {author}
  {\bibfnamefont {A.~R.}\ \bibnamefont {Johnson}}, \bibinfo {author}
  {\bibfnamefont {Y.}~\bibnamefont {Okawachi}}, \bibinfo {author}
  {\bibfnamefont {M.~R.~E.}\ \bibnamefont {Lamont}}, \bibinfo {author}
  {\bibfnamefont {M.}~\bibnamefont {Lipson}},\ and\ \bibinfo {author}
  {\bibfnamefont {A.~L.}\ \bibnamefont {Gaeta}},\ }\bibfield  {title} {\bibinfo
  {title} {Strong polarization mode coupling in microresonators},\ }\href
  {https://doi.org/10.1364/OL.39.005134} {\bibfield  {journal} {\bibinfo
  {journal} {Opt. Lett.}\ }\textbf {\bibinfo {volume} {39}},\ \bibinfo {pages}
  {5134} (\bibinfo {year} {2014})}\BibitemShut {NoStop}%
\bibitem [{\citenamefont {Mazzei}\ \emph {et~al.}(2007)\citenamefont {Mazzei},
  \citenamefont {G\"otzinger}, \citenamefont {de~S.~Menezes}, \citenamefont
  {Zumofen}, \citenamefont {Benson},\ and\ \citenamefont
  {Sandoghdar}}]{PhysRevLett.99.173603}%
  \BibitemOpen
  \bibfield  {author} {\bibinfo {author} {\bibfnamefont {A.}~\bibnamefont
  {Mazzei}}, \bibinfo {author} {\bibfnamefont {S.}~\bibnamefont {G\"otzinger}},
  \bibinfo {author} {\bibfnamefont {L.}~\bibnamefont {de~S.~Menezes}}, \bibinfo
  {author} {\bibfnamefont {G.}~\bibnamefont {Zumofen}}, \bibinfo {author}
  {\bibfnamefont {O.}~\bibnamefont {Benson}},\ and\ \bibinfo {author}
  {\bibfnamefont {V.}~\bibnamefont {Sandoghdar}},\ }\bibfield  {title}
  {\bibinfo {title} {Controlled coupling of counterpropagating
  whispering-gallery modes by a single {R}ayleigh scatterer: {A} classical
  problem in a quantum optical light},\ }\href
  {https://doi.org/10.1103/PhysRevLett.99.173603} {\bibfield  {journal}
  {\bibinfo  {journal} {Phys. Rev. Lett.}\ }\textbf {\bibinfo {volume} {99}},\
  \bibinfo {pages} {173603} (\bibinfo {year} {2007})}\BibitemShut {NoStop}%
\bibitem [{\citenamefont {Yoshiki}\ \emph {et~al.}(2015)\citenamefont
  {Yoshiki}, \citenamefont {Chen-Jinnai}, \citenamefont {Tetsumoto},\ and\
  \citenamefont {Tanabe}}]{Yoshiki:15}%
  \BibitemOpen
  \bibfield  {author} {\bibinfo {author} {\bibfnamefont {W.}~\bibnamefont
  {Yoshiki}}, \bibinfo {author} {\bibfnamefont {A.}~\bibnamefont
  {Chen-Jinnai}}, \bibinfo {author} {\bibfnamefont {T.}~\bibnamefont
  {Tetsumoto}},\ and\ \bibinfo {author} {\bibfnamefont {T.}~\bibnamefont
  {Tanabe}},\ }\bibfield  {title} {\bibinfo {title} {Observation of energy
  oscillation between strongly-coupled counter-propagating ultra-high {Q}
  whispering gallery modes},\ }\href {https://doi.org/10.1364/OE.23.030851}
  {\bibfield  {journal} {\bibinfo  {journal} {Opt. Express}\ }\textbf {\bibinfo
  {volume} {23}},\ \bibinfo {pages} {30851} (\bibinfo {year}
  {2015})}\BibitemShut {NoStop}%
\bibitem [{\citenamefont {Fujii}\ \emph {et~al.}(2017)\citenamefont {Fujii},
  \citenamefont {Hori}, \citenamefont {Kato}, \citenamefont {Suzuki},
  \citenamefont {Okabe}, \citenamefont {Yoshiki}, \citenamefont {Jinnai},\ and\
  \citenamefont {Tanabe}}]{Fujii:17}%
  \BibitemOpen
  \bibfield  {author} {\bibinfo {author} {\bibfnamefont {S.}~\bibnamefont
  {Fujii}}, \bibinfo {author} {\bibfnamefont {A.}~\bibnamefont {Hori}},
  \bibinfo {author} {\bibfnamefont {T.}~\bibnamefont {Kato}}, \bibinfo {author}
  {\bibfnamefont {R.}~\bibnamefont {Suzuki}}, \bibinfo {author} {\bibfnamefont
  {Y.}~\bibnamefont {Okabe}}, \bibinfo {author} {\bibfnamefont
  {W.}~\bibnamefont {Yoshiki}}, \bibinfo {author} {\bibfnamefont {A.-C.}\
  \bibnamefont {Jinnai}},\ and\ \bibinfo {author} {\bibfnamefont
  {T.}~\bibnamefont {Tanabe}},\ }\bibfield  {title} {\bibinfo {title} {Effect
  on {K}err comb generation in a clockwise and counter-clockwise mode coupled
  microcavity},\ }\href {https://doi.org/10.1364/OE.25.028969} {\bibfield
  {journal} {\bibinfo  {journal} {Opt. Express}\ }\textbf {\bibinfo {volume}
  {25}},\ \bibinfo {pages} {28969} (\bibinfo {year} {2017})}\BibitemShut
  {NoStop}%
\bibitem [{\citenamefont {Yang}\ \emph {et~al.}(2017)\citenamefont {Yang},
  \citenamefont {Yi},\ and\ \citenamefont {Vahala}}]{Vahala_18np}%
  \BibitemOpen
  \bibfield  {author} {\bibinfo {author} {\bibfnamefont {Q.-F.}\ \bibnamefont
  {Yang}}, \bibinfo {author} {\bibfnamefont {X.}~\bibnamefont {Yi}},\ and\
  \bibinfo {author} {\bibfnamefont {K.}~\bibnamefont {Vahala}},\ }\bibfield
  {title} {\bibinfo {title} {Counter-propagating solitons in microresonators},\
  }\href {https://doi.org/10.1038/nphoton.2017.117} {\bibfield  {journal}
  {\bibinfo  {journal} {Nat. Photon.}\ }\textbf {\bibinfo {volume} {11}},\
  \bibinfo {pages} {560} (\bibinfo {year} {2017})}\BibitemShut {NoStop}%
\bibitem [{\citenamefont {Joshi}\ \emph {et~al.}(2018)\citenamefont {Joshi},
  \citenamefont {Klenner}, \citenamefont {Okawachi}, \citenamefont {Yu},
  \citenamefont {Luke}, \citenamefont {Ji}, \citenamefont {Lipson},\ and\
  \citenamefont {Gaeta}}]{Joshi:18}%
  \BibitemOpen
  \bibfield  {author} {\bibinfo {author} {\bibfnamefont {C.}~\bibnamefont
  {Joshi}}, \bibinfo {author} {\bibfnamefont {A.}~\bibnamefont {Klenner}},
  \bibinfo {author} {\bibfnamefont {Y.}~\bibnamefont {Okawachi}}, \bibinfo
  {author} {\bibfnamefont {M.}~\bibnamefont {Yu}}, \bibinfo {author}
  {\bibfnamefont {K.}~\bibnamefont {Luke}}, \bibinfo {author} {\bibfnamefont
  {X.}~\bibnamefont {Ji}}, \bibinfo {author} {\bibfnamefont {M.}~\bibnamefont
  {Lipson}},\ and\ \bibinfo {author} {\bibfnamefont {A.~L.}\ \bibnamefont
  {Gaeta}},\ }\bibfield  {title} {\bibinfo {title} {Counter-rotating cavity
  solitons in a silicon nitride microresonator},\ }\href
  {https://doi.org/10.1364/OL.43.000547} {\bibfield  {journal} {\bibinfo
  {journal} {Opt. Lett.}\ }\textbf {\bibinfo {volume} {43}},\ \bibinfo {pages}
  {547} (\bibinfo {year} {2018})}\BibitemShut {NoStop}%
\bibitem [{\citenamefont {Nielsen}\ \emph {et~al.}(2019)\citenamefont
  {Nielsen}, \citenamefont {Garbin}, \citenamefont {Coen}, \citenamefont
  {Murdoch},\ and\ \citenamefont {Erkintalo}}]{Nielsen2019}%
  \BibitemOpen
  \bibfield  {author} {\bibinfo {author} {\bibfnamefont {A.~U.}\ \bibnamefont
  {Nielsen}}, \bibinfo {author} {\bibfnamefont {B.}~\bibnamefont {Garbin}},
  \bibinfo {author} {\bibfnamefont {S.}~\bibnamefont {Coen}}, \bibinfo {author}
  {\bibfnamefont {S.~G.}\ \bibnamefont {Murdoch}},\ and\ \bibinfo {author}
  {\bibfnamefont {M.}~\bibnamefont {Erkintalo}},\ }\bibfield  {title} {\bibinfo
  {title} {Coexistence and interactions between nonlinear states with different
  polarizations in a monochromatically driven passive kerr resonator},\ }\href
  {https://doi.org/10.1103/PhysRevLett.123.013902} {\bibfield  {journal}
  {\bibinfo  {journal} {Phys. Rev. Lett.}\ }\textbf {\bibinfo {volume} {123}},\
  \bibinfo {pages} {013902} (\bibinfo {year} {2019})}\BibitemShut {NoStop}%
\bibitem [{\citenamefont {Boyd}(2013)}]{boyd2013nonlinear}%
  \BibitemOpen
  \bibfield  {author} {\bibinfo {author} {\bibfnamefont {R.}~\bibnamefont
  {Boyd}},\ }\href {https://books.google.ru/books?id=\_YpGBQAAQBAJ} {\emph
  {\bibinfo {title} {Nonlinear Optics}}}\ (\bibinfo  {publisher} {Elsevier
  Science},\ \bibinfo {year} {2013})\BibitemShut {NoStop}%
\bibitem [{\citenamefont {Gorodetsky}\ and\ \citenamefont
  {Ilchenko}(1999)}]{Gorodetsky:99}%
  \BibitemOpen
  \bibfield  {author} {\bibinfo {author} {\bibfnamefont {M.~L.}\ \bibnamefont
  {Gorodetsky}}\ and\ \bibinfo {author} {\bibfnamefont {V.~S.}\ \bibnamefont
  {Ilchenko}},\ }\bibfield  {title} {\bibinfo {title} {Optical microsphere
  resonators: optimal coupling to high-{Q} whispering-gallery modes},\ }\href
  {https://doi.org/10.1364/JOSAB.16.000147} {\bibfield  {journal} {\bibinfo
  {journal} {J. Opt. Soc. Am. B}\ }\textbf {\bibinfo {volume} {16}},\ \bibinfo
  {pages} {147} (\bibinfo {year} {1999})}\BibitemShut {NoStop}%
\bibitem [{\citenamefont {Deych}(2011)}]{Deych2011}%
  \BibitemOpen
  \bibfield  {author} {\bibinfo {author} {\bibfnamefont {L.}~\bibnamefont
  {Deych}},\ }\bibfield  {title} {\bibinfo {title} {Comment on ``{M}odal
  expansion approach to optical-frequency-comb generation with monolithic
  whispering-gallery-mode resonators''},\ }\href
  {https://doi.org/10.1103/PhysRevA.84.017801} {\bibfield  {journal} {\bibinfo
  {journal} {Phys. Rev. A}\ }\textbf {\bibinfo {volume} {84}},\ \bibinfo
  {pages} {017801} (\bibinfo {year} {2011})}\BibitemShut {NoStop}%
\bibitem [{\citenamefont {Lai}\ \emph {et~al.}(1990)\citenamefont {Lai},
  \citenamefont {Leung}, \citenamefont {Young}, \citenamefont {Barber},\ and\
  \citenamefont {Hill}}]{Hill1990}%
  \BibitemOpen
  \bibfield  {author} {\bibinfo {author} {\bibfnamefont {H.~M.}\ \bibnamefont
  {Lai}}, \bibinfo {author} {\bibfnamefont {P.~T.}\ \bibnamefont {Leung}},
  \bibinfo {author} {\bibfnamefont {K.}~\bibnamefont {Young}}, \bibinfo
  {author} {\bibfnamefont {P.~W.}\ \bibnamefont {Barber}},\ and\ \bibinfo
  {author} {\bibfnamefont {S.~C.}\ \bibnamefont {Hill}},\ }\bibfield  {title}
  {\bibinfo {title} {Time-independent perturbation for leaking electromagnetic
  modes in open systems with application to resonances in microdroplets},\
  }\href {https://doi.org/10.1103/PhysRevA.41.5187} {\bibfield  {journal}
  {\bibinfo  {journal} {Phys. Rev. A}\ }\textbf {\bibinfo {volume} {41}},\
  \bibinfo {pages} {5187} (\bibinfo {year} {1990})}\BibitemShut {NoStop}%
\bibitem [{\citenamefont {Chembo}\ and\ \citenamefont {Yu}(2010)}]{Chembo2010}%
  \BibitemOpen
  \bibfield  {author} {\bibinfo {author} {\bibfnamefont {Y.~K.}\ \bibnamefont
  {Chembo}}\ and\ \bibinfo {author} {\bibfnamefont {N.}~\bibnamefont {Yu}},\
  }\bibfield  {title} {\bibinfo {title} {Modal expansion approach to
  optical-frequency-comb generation with monolithic whispering-gallery-mode
  resonators},\ }\href {https://doi.org/10.1103/PhysRevA.82.033801} {\bibfield
  {journal} {\bibinfo  {journal} {Phys. Rev. A}\ }\textbf {\bibinfo {volume}
  {82}},\ \bibinfo {pages} {033801} (\bibinfo {year} {2010})}\BibitemShut
  {NoStop}%
\bibitem [{\citenamefont {Cherenkov}\ \emph {et~al.}(2017)\citenamefont
  {Cherenkov}, \citenamefont {Kondratiev}, \citenamefont {Lobanov},
  \citenamefont {Shitikov}, \citenamefont {Skryabin},\ and\ \citenamefont
  {Gorodetsky}}]{Cherenkov:17}%
  \BibitemOpen
  \bibfield  {author} {\bibinfo {author} {\bibfnamefont {A.~V.}\ \bibnamefont
  {Cherenkov}}, \bibinfo {author} {\bibfnamefont {N.~M.}\ \bibnamefont
  {Kondratiev}}, \bibinfo {author} {\bibfnamefont {V.~E.}\ \bibnamefont
  {Lobanov}}, \bibinfo {author} {\bibfnamefont {A.~E.}\ \bibnamefont
  {Shitikov}}, \bibinfo {author} {\bibfnamefont {D.~V.}\ \bibnamefont
  {Skryabin}},\ and\ \bibinfo {author} {\bibfnamefont {M.~L.}\ \bibnamefont
  {Gorodetsky}},\ }\bibfield  {title} {\bibinfo {title} {Raman-kerr frequency
  combs in microresonators with normal dispersion},\ }\href
  {https://doi.org/10.1364/OE.25.031148} {\bibfield  {journal} {\bibinfo
  {journal} {Opt. Express}\ }\textbf {\bibinfo {volume} {25}},\ \bibinfo
  {pages} {31148} (\bibinfo {year} {2017})}\BibitemShut {NoStop}%
\bibitem [{\citenamefont {Zhu}\ \emph {et~al.}(2010)\citenamefont {Zhu},
  \citenamefont {Ozdemir}, \citenamefont {Xiao}, \citenamefont {Li},
  \citenamefont {He}, \citenamefont {Chen},\ and\ \citenamefont
  {Yang}}]{Zhu2010}%
  \BibitemOpen
  \bibfield  {author} {\bibinfo {author} {\bibfnamefont {J.}~\bibnamefont
  {Zhu}}, \bibinfo {author} {\bibfnamefont {S.~K.}\ \bibnamefont {Ozdemir}},
  \bibinfo {author} {\bibfnamefont {Y.-F.}\ \bibnamefont {Xiao}}, \bibinfo
  {author} {\bibfnamefont {L.}~\bibnamefont {Li}}, \bibinfo {author}
  {\bibfnamefont {L.}~\bibnamefont {He}}, \bibinfo {author} {\bibfnamefont
  {D.-R.}\ \bibnamefont {Chen}},\ and\ \bibinfo {author} {\bibfnamefont
  {L.}~\bibnamefont {Yang}},\ }\bibfield  {title} {\bibinfo {title} {On-chip
  single nanoparticle detection and sizing by mode splitting in an ultrahigh-q
  microresonator},\ }\href {https://doi.org/10.1038/nphoton.2009.237}
  {\bibfield  {journal} {\bibinfo  {journal} {Nature Photonics}\ }\textbf
  {\bibinfo {volume} {4}},\ \bibinfo {pages} {46} (\bibinfo {year}
  {2010})}\BibitemShut {NoStop}%
\bibitem [{\citenamefont {Li}\ \emph {et~al.}(2012)\citenamefont {Li},
  \citenamefont {Eftekhar}, \citenamefont {Xia},\ and\ \citenamefont
  {Adibi}}]{Li:12}%
  \BibitemOpen
  \bibfield  {author} {\bibinfo {author} {\bibfnamefont {Q.}~\bibnamefont
  {Li}}, \bibinfo {author} {\bibfnamefont {A.~A.}\ \bibnamefont {Eftekhar}},
  \bibinfo {author} {\bibfnamefont {Z.}~\bibnamefont {Xia}},\ and\ \bibinfo
  {author} {\bibfnamefont {A.}~\bibnamefont {Adibi}},\ }\bibfield  {title}
  {\bibinfo {title} {Azimuthal-order variations of surface-roughness-induced
  mode splitting and scattering loss in high-q microdisk resonators},\ }\href
  {https://doi.org/10.1364/OL.37.001586} {\bibfield  {journal} {\bibinfo
  {journal} {Opt. Lett.}\ }\textbf {\bibinfo {volume} {37}},\ \bibinfo {pages}
  {1586} (\bibinfo {year} {2012})}\BibitemShut {NoStop}%
\bibitem [{\citenamefont {Hansson}\ \emph {et~al.}(2014)\citenamefont
  {Hansson}, \citenamefont {Modotto},\ and\ \citenamefont
  {Wabnitz}}]{HANSSON2014134}%
  \BibitemOpen
  \bibfield  {author} {\bibinfo {author} {\bibfnamefont {T.}~\bibnamefont
  {Hansson}}, \bibinfo {author} {\bibfnamefont {D.}~\bibnamefont {Modotto}},\
  and\ \bibinfo {author} {\bibfnamefont {S.}~\bibnamefont {Wabnitz}},\
  }\bibfield  {title} {\bibinfo {title} {On the numerical simulation of {K}err
  frequency combs using coupled mode equations},\ }\href
  {https://doi.org/https://doi.org/10.1016/j.optcom.2013.09.017} {\bibfield
  {journal} {\bibinfo  {journal} {Optics Communications}\ }\textbf {\bibinfo
  {volume} {312}},\ \bibinfo {pages} {134 } (\bibinfo {year}
  {2014})}\BibitemShut {NoStop}%
\bibitem [{\citenamefont {Lugiato}\ and\ \citenamefont
  {Lefever}(1987)}]{PhysRevLett.58.2209}%
  \BibitemOpen
  \bibfield  {author} {\bibinfo {author} {\bibfnamefont {L.~A.}\ \bibnamefont
  {Lugiato}}\ and\ \bibinfo {author} {\bibfnamefont {R.}~\bibnamefont
  {Lefever}},\ }\bibfield  {title} {\bibinfo {title} {Spatial dissipative
  structures in passive optical systems},\ }\href
  {https://doi.org/10.1103/PhysRevLett.58.2209} {\bibfield  {journal} {\bibinfo
   {journal} {Phys. Rev. Lett.}\ }\textbf {\bibinfo {volume} {58}},\ \bibinfo
  {pages} {2209} (\bibinfo {year} {1987})}\BibitemShut {NoStop}%
\bibitem [{\citenamefont {Chembo}\ and\ \citenamefont
  {Menyuk}(2013)}]{PhysRevA.87.053852}%
  \BibitemOpen
  \bibfield  {author} {\bibinfo {author} {\bibfnamefont {Y.~K.}\ \bibnamefont
  {Chembo}}\ and\ \bibinfo {author} {\bibfnamefont {C.~R.}\ \bibnamefont
  {Menyuk}},\ }\bibfield  {title} {\bibinfo {title} {Spatiotemporal
  {L}ugiato-{L}efever formalism for {K}err-comb generation in
  whispering-gallery-mode resonators},\ }\href
  {https://doi.org/10.1103/PhysRevA.87.053852} {\bibfield  {journal} {\bibinfo
  {journal} {Phys. Rev. A}\ }\textbf {\bibinfo {volume} {87}},\ \bibinfo
  {pages} {053852} (\bibinfo {year} {2013})}\BibitemShut {NoStop}%
\bibitem [{\citenamefont {Lugiato}\ \emph {et~al.}(2018)\citenamefont
  {Lugiato}, \citenamefont {Prati}, \citenamefont {Gorodetsky},\ and\
  \citenamefont {Kippenberg}}]{doi:10.1098/rsta.2018.0113}%
  \BibitemOpen
  \bibfield  {author} {\bibinfo {author} {\bibfnamefont {L.~A.}\ \bibnamefont
  {Lugiato}}, \bibinfo {author} {\bibfnamefont {F.}~\bibnamefont {Prati}},
  \bibinfo {author} {\bibfnamefont {M.~L.}\ \bibnamefont {Gorodetsky}},\ and\
  \bibinfo {author} {\bibfnamefont {T.~J.}\ \bibnamefont {Kippenberg}},\
  }\bibfield  {title} {\bibinfo {title} {From the {L}ugiato-{L}efever equation
  to microresonator-based soliton {K}err frequency combs},\ }\href
  {https://doi.org/10.1098/rsta.2018.0113} {\bibfield  {journal} {\bibinfo
  {journal} {Philosophical Transactions of the Royal Society A: Mathematical,
  Physical and Engineering Sciences}\ }\textbf {\bibinfo {volume} {376}},\
  \bibinfo {pages} {20180113} (\bibinfo {year} {2018})}\BibitemShut {NoStop}%
\bibitem [{\citenamefont {Weiss}\ \emph {et~al.}(1995)\citenamefont {Weiss},
  \citenamefont {Sandoghdar}, \citenamefont {Hare}, \citenamefont
  {Lef\`{e}vre-Seguin}, \citenamefont {Raimond},\ and\ \citenamefont
  {Haroche}}]{Weiss:95}%
  \BibitemOpen
  \bibfield  {author} {\bibinfo {author} {\bibfnamefont {D.~S.}\ \bibnamefont
  {Weiss}}, \bibinfo {author} {\bibfnamefont {V.}~\bibnamefont {Sandoghdar}},
  \bibinfo {author} {\bibfnamefont {J.}~\bibnamefont {Hare}}, \bibinfo {author}
  {\bibfnamefont {V.}~\bibnamefont {Lef\`{e}vre-Seguin}}, \bibinfo {author}
  {\bibfnamefont {J.-M.}\ \bibnamefont {Raimond}},\ and\ \bibinfo {author}
  {\bibfnamefont {S.}~\bibnamefont {Haroche}},\ }\bibfield  {title} {\bibinfo
  {title} {Splitting of high-{Q} {M}ie modes induced by light backscattering in
  silica microspheres},\ }\href {https://doi.org/10.1364/OL.20.001835}
  {\bibfield  {journal} {\bibinfo  {journal} {Opt. Lett.}\ }\textbf {\bibinfo
  {volume} {20}},\ \bibinfo {pages} {1835} (\bibinfo {year}
  {1995})}\BibitemShut {NoStop}%
\bibitem [{\citenamefont {Kippenberg}\ \emph {et~al.}(2002)\citenamefont
  {Kippenberg}, \citenamefont {Spillane},\ and\ \citenamefont
  {Vahala}}]{Kippenberg:02}%
  \BibitemOpen
  \bibfield  {author} {\bibinfo {author} {\bibfnamefont {T.~J.}\ \bibnamefont
  {Kippenberg}}, \bibinfo {author} {\bibfnamefont {S.~M.}\ \bibnamefont
  {Spillane}},\ and\ \bibinfo {author} {\bibfnamefont {K.~J.}\ \bibnamefont
  {Vahala}},\ }\bibfield  {title} {\bibinfo {title} {Modal coupling in
  traveling-wave resonators},\ }\href {https://doi.org/10.1364/OL.27.001669}
  {\bibfield  {journal} {\bibinfo  {journal} {Opt. Lett.}\ }\textbf {\bibinfo
  {volume} {27}},\ \bibinfo {pages} {1669} (\bibinfo {year}
  {2002})}\BibitemShut {NoStop}%
\bibitem [{\citenamefont {Savchenkov}\ \emph {et~al.}(2007)\citenamefont
  {Savchenkov}, \citenamefont {Matsko}, \citenamefont {Ilchenko},\ and\
  \citenamefont {Maleki}}]{Savchenkov:07}%
  \BibitemOpen
  \bibfield  {author} {\bibinfo {author} {\bibfnamefont {A.~A.}\ \bibnamefont
  {Savchenkov}}, \bibinfo {author} {\bibfnamefont {A.~B.}\ \bibnamefont
  {Matsko}}, \bibinfo {author} {\bibfnamefont {V.~S.}\ \bibnamefont
  {Ilchenko}},\ and\ \bibinfo {author} {\bibfnamefont {L.}~\bibnamefont
  {Maleki}},\ }\bibfield  {title} {\bibinfo {title} {Optical resonators with
  ten million finesse},\ }\href {https://doi.org/10.1364/OE.15.006768}
  {\bibfield  {journal} {\bibinfo  {journal} {Opt. Express}\ }\textbf {\bibinfo
  {volume} {15}},\ \bibinfo {pages} {6768} (\bibinfo {year}
  {2007})}\BibitemShut {NoStop}%
\bibitem [{\citenamefont {Lobanov}\ \emph {et~al.}(2015)\citenamefont
  {Lobanov}, \citenamefont {Lihachev}, \citenamefont {Kippenberg},\ and\
  \citenamefont {Gorodetsky}}]{Lobanov:15}%
  \BibitemOpen
  \bibfield  {author} {\bibinfo {author} {\bibfnamefont {V.}~\bibnamefont
  {Lobanov}}, \bibinfo {author} {\bibfnamefont {G.}~\bibnamefont {Lihachev}},
  \bibinfo {author} {\bibfnamefont {T.~J.}\ \bibnamefont {Kippenberg}},\ and\
  \bibinfo {author} {\bibfnamefont {M.}~\bibnamefont {Gorodetsky}},\ }\bibfield
   {title} {\bibinfo {title} {Frequency combs and platicons in optical
  microresonators with normal {GVD}},\ }\href
  {https://doi.org/10.1364/OE.23.007713} {\bibfield  {journal} {\bibinfo
  {journal} {Opt. Express}\ }\textbf {\bibinfo {volume} {23}},\ \bibinfo
  {pages} {7713} (\bibinfo {year} {2015})}\BibitemShut {NoStop}%
\bibitem [{\citenamefont {Lobanov}\ \emph {et~al.}(2017)\citenamefont
  {Lobanov}, \citenamefont {Cherenkov}, \citenamefont {Shitikov}, \citenamefont
  {Bilenko},\ and\ \citenamefont {Gorodetsky}}]{Lobanov2017}%
  \BibitemOpen
  \bibfield  {author} {\bibinfo {author} {\bibfnamefont {V.~E.}\ \bibnamefont
  {Lobanov}}, \bibinfo {author} {\bibfnamefont {A.~V.}\ \bibnamefont
  {Cherenkov}}, \bibinfo {author} {\bibfnamefont {A.~E.}\ \bibnamefont
  {Shitikov}}, \bibinfo {author} {\bibfnamefont {I.~A.}\ \bibnamefont
  {Bilenko}},\ and\ \bibinfo {author} {\bibfnamefont {M.~L.}\ \bibnamefont
  {Gorodetsky}},\ }\bibfield  {title} {\bibinfo {title} {Dynamics of platicons
  due to third-order dispersion},\ }\href
  {https://doi.org/10.1140/epjd/e2017-80148-0} {\bibfield  {journal} {\bibinfo
  {journal} {The European Physical Journal D}\ }\textbf {\bibinfo {volume}
  {71}},\ \bibinfo {pages} {185} (\bibinfo {year} {2017})}\BibitemShut
  {NoStop}%
\end{thebibliography}%

\end{document}